\documentclass{aa}
\makeatletter
  \RequirePackage{silence}

\makeatother

\makeatletter
  \RequirePackage{silence}
\makeatother

\makeatletter
  \RequirePackage{silence}

\makeatother

\usepackage{graphicx}

\usepackage{txfonts}

\usepackage[pdfpagelabels=false]{hyperref}
\usepackage{tabularx}
\usepackage{subcaption}
\usepackage{float}
\usepackage{placeins}
\usepackage{silence}
\usepackage{natbib}

\usepackage[version=4]{mhchem}
\usepackage{siunitx}
\usepackage{caption}

\FloatBarrier

\hypersetup{colorlinks=true,linkcolor=blue,citecolor=blue,filecolor=blue,urlcolor=blue,}

\newcommand{\vsys}{$v_{\text{sys}}$}
\newcommand{\vsini}{$v_{\mathrm{rot}}\sin{i}$}
\newenvironment{appendices}{\begin{appendix}}{\end{appendix}}

\begin{document}

\title{Sorting a mess}
\subtitle{I. Addressing velocity-axis correlation of cross-correlation functions. A census of atomic and ionised species in KELT-9b's atmosphere}
\titlerunning{Sorting a mess. I.}

\author{N. W. Borsato\inst{1,2,3,4}
\and
B. Thorsbro\inst{2,5}
\and
H. J. Hoeijmakers\inst{2}
\and
D. B. Zucker\inst{3,4}
\and
S. Pelletier\inst{6}
\and
B. Prinoth\inst{2}
}

\institute{
Univ Toulouse, CNES, CNRS, IRAP, 14 avenue Edouard Belin, 31400 Toulouse, France\\
\email{nicholas.borsato@univ-tlse3.fr}
\and
Lund Observatory, Division of Astronomy and Theoretical Physics, Lund University, Box 43, 221 00 Lund, Sweden
\and
School of Mathematical and Physical Sciences, Macquarie University, Sydney, NSW 2109, Australia
\and
Astrophysics and Space Technologies Research Centre, Macquarie University, Sydney, NSW 2109, Australia
\and
Observatoire de la Côte d’Azur, Laboratoire Lagrange, CNRS, Université Côte d’Azur, Blvd de l’Observatoire, 06304 Nice, France
\and
Observatoire astronomique de l'Université de Genève, 51 chemin Pegasi 1290 Versoix, Switzerland
}

\date{Received 4 August 2025; accepted 1 September 2026}

\abstract
{
Ultra-hot Jupiters (UHJs) represent the most extreme class of close-in gas giants, with KELT-9b being the most extreme example known. In this study, we combine 13 transit observations of KELT-9b from six high-resolution spectrographs spanning 316--960\,nm to conduct a census of atomic and ionised species in the planet’s atmosphere using the cross-correlation technique. The primary objective is to introduce an agnostic strategy to mitigate velocity-axis correlation bias in parameter uncertainties during cross-correlation analysis. By employing a forward model of the cross-correlation function (CCF) alongside a Monte Carlo bootstrap of flux uncertainties, the individual CCF-derived velocity profiles of each species in the planet’s atmosphere are measured, and potential time dependence in these velocity profiles is assessed. The stellar reference frame is determined, yielding a systemic velocity of $V_{\mathrm{sys}} = -17.99 \pm 0.04$\,km\,s$^{-1}$ and a projected rotational velocity of $v\sin i = 110.54 \pm 0.05$\,km\,s$^{-1}$, consistent across all instruments and observing nights. The combined planetary absorption profiles provide insight into the collective motion of all detectable gas species in the atmosphere. All previously reported species are recovered except Tb\,II, and the combined analysis detects 29 species, including additional absorbers C\,I, Si\,I, K\,I, V\,II, Co\,I, Zr\,II, La\,II, Pr\,II, and Nd\,II. All 29 recovered species exceed the conventional $5\sigma$ threshold in the bootstrap Mahalanobis significance test. The collective Doppler shifts display a blueshift of $\Delta V_{\mathrm{offset}} = -8.02 \pm 0.17$\,km\,s$^{-1}$ relative to the stellar rest frame, indicating a net day-to-night wind, with no explicit time dependence observed when all species are homogenised. These results constitute a catalogue of detections and retrieved Doppler parameters, forming a foundation for a companion study of KELT-9b that will analyse the most significant detections and compare differences across individual species.
}


\maketitle
\section{Introduction}
Ultra--hot Jupiters (UHJs) are an extreme class of exoplanet, orbiting so close to their host stars that their dayside temperatures exceed 2200\,K~\citep{Parmentier_2018}. In this temperature regime, molecular hydrogen becomes dissociated, resulting in an atmosphere containing a high abundance of atomic and ionised species~\citep{Lothringer_2018,Parmentier_2018}, which generally absorb strongly in the optical and are accessible even with modestly sized telescopes~\citep{Bello_Arufe_2022,Lowson_2023,Borsato_2024}. Additionally, UHJs occupy a large swathe of atmospheric processes that we can observe and study, such as cloud formation~\citep[e.g.][]{Helling_2021}, cold-trapping~\citep[e.g.][]{Spiegel_2009}, and temperature inversions~\citep[e.g.][]{Pino_2020,May_2021}, acting as convenient laboratories for exoplanet atmospheric studies.

Previous high-resolution spectroscopic studies of ultra-hot Jupiters have provided direct evidence for a range of atmospheric processes. Observations of refractory elements such as titanium and iron have revealed temperature-dependent cold-trapping, with depleted abundances in cooler systems and persistent detections in the hottest cases \citep[e.g.][]{Gandhi_2023, Pelletier_2023,pelletier_coldtrap_2026,Hoeijmakers_2024}. Doppler-shifted line profiles have been used to measure high-altitude winds and day--night circulation patterns \citep[e.g.][]{Landman_2021,Asnodkar_2022a,Stangret_2024,Langeveld_2025}. Detections of neutral and ionised species constrain temperatures within different regions of the planet's atmosphere \citep{Prinoth_2023}, while detections of extended absorption features beyond the Roche lobe indicate ongoing atmospheric escape \citep[e.g.][]{Turner_2020,Lowson_2023,Czesla_2024a,Czesla_2024b}. Differences in species detected at varying pressures have also been used to infer vertical mixing and stratification \citep{Seidel_2025}. Phase-resolved high-resolution spectroscopy has further revealed that atmospheric absorption signals can evolve throughout transit, suggesting complex three-dimensional structures and spatially varying dynamics \citep{Ehrenreich_2020,Prinoth_2023,Langeveld_2025}. Together, these results demonstrate the diagnostic power of high-resolution spectroscopy for probing the chemical, thermal, and dynamical properties of ultra-hot Jupiter atmospheres.

Many of the atmospheric inferences drawn from hot giant exoplanets have come from transmission spectroscopy, which probes starlight transmitted through the terminator region during transit. This technique has proven particularly effective for ultra-hot Jupiters, where extended, atomic-rich atmospheres imprint strong absorption signatures on the stellar spectrum. These detections yield not only a census of atmospheric species but also a pathway to constraining elemental abundances, which in turn inform models of planet formation and migration \citep{Line_2021,Pelletier_2021,Pelletier_2025,Smith_2024}. While complementary insights have been obtained through dayside emission spectroscopy \citep[e.g.][]{Brogi_2012,de_Kok_2013,Yan_2020,Pino_2020,Nugroho_2020,Kasper_2021,Borsa_2022,Yan_2022,Yan_2023,Ridden_Harper_2023,Brogi_2023,Deibert_2024,Guo_2024}, transmission spectroscopy remains the dominant method for detecting species and characterising upper-atmospheric structure \citep[e.g.][]{Hoeijmakers_2018,Cauley_2019,Yan_2019,Hoeijmakers_2019,Casasayas_Barris_2019,Hoeijmakers_2020,Cabot_2020,Ben_Yami_2020,Yan_2021,Landman_2021,Merritt_2021,Sanchez_Lopez_2022,Borsa_2022b_OIdetection}. Among the tools used, the cross-correlation technique has played a central role. By coherently combining hundreds to thousands of spectral lines into a single velocity-resolved signal, the cross-correlation technique enables the detection of species whose individual lines fall below the noise level of the spectrum. This method has also been used to infer wind speeds, orbital dynamics, and thermospheric properties \citep{Snellen_2010_CC}, and has become a cornerstone of high-resolution studies of exoplanet atmospheres.

One of the key limitations of the cross-correlation function (CCF) is that there are underlying problems with the method that can restrict analysis of the results. An example of such a problem is aliases, where blended lines from other strongly absorbing species introduce additional absorption features which can have high statistical significance~\citep{Borsato_2023}. Additionally, correlated noise and sampling artefacts can mimic real detections \citep{Birkby_2018}. Moreover, the cross-correlation process itself breaks the assumption of independent noise, introducing velocity-axis correlation in the data which complicates robust error estimation~\citep{Hoeijmakers_2019}. These effects can inflate detection significances and hinder efforts to extract physical parameters, often limiting analyses to qualitative interpretations rather than statistically rigorous results. If velocity-axis correlation in the data can be dealt with, however, the CCF could become a tool to analyse CCF-derived velocity profiles of individual species detected at high significance, with stronger confidence that the uncertainties are unbiased. Differences in these values could then start to become linked to the atmospheric structure of the planet, as has been suggested in previous studies~\citep{Prinoth_2025,Seidel_2025}.

This study therefore attempts to address the issues of velocity-axis correlation, by developing a methodology that propagates the uncertainties in the data in a statistically consistent manner, and attempts to apply that framework by combining 13 transit observations of the ultra-hot Jupiter KELT-9b from six different high-resolution spectrographs to build a comprehensive dataset spanning 316–960 nm. As the hottest known exoplanet orbiting a main-sequence A-type star, with an equilibrium temperature of approximately 4000\,K~\citep{Gaudi_2017}, its atmosphere is dominated by atomic and ionised species, with expected minimal cloud formation or cold-trapping to obscure spectroscopic signals~\citep{Kitzmann_2018}. Its extreme conditions make it both an excellent laboratory for atmospheric characterisation and to test our velocity-axis correlation mitigation strategy. This paper is the first in a two-part series, to develop the methodology and perform a census of KELT-9b’s atmospheric gas species, and to present the global results of our cross-correlation attempts. Future work will analyse the detections on a species-by-species basis in an attempt to link their differences to the dynamical behaviour and geometrical structure of KELT-9b’s atmosphere.

\section{Observations and data}\label{sect:observations_and_data}
This study combines 13 transit observations of KELT-9b obtained with six high-resolution spectrographs: HARPS-N \citep{Mayor_2003}, CARMENES \citep{Quirrenbach_2016_CARMENES}, FIES \citep{Telting_2014_FIES}, FOCES \citep{Pfeiffer_1998, Wang_et_al_2017}, HIRES \citep{Vogt_1994_HIRES_Spectrograph}, and MAROON-X \citep{Seifahrt_2018_MAROONX, Seifahrt_2020_MAROONX}, spanning a wavelength range of 316--960~nm. Three transits were observed by our team: one with HIRES on Keck (AS22B-073, principal investigator (PI): Thorsbro), and two with FIES on the Nordic Optical Telescope (65-001, PI: Borsato). All spectra were reduced using the standard, instrument-specific pipelines described in the literature. The remaining observations were obtained from archival datasets. In this section, we discuss the characteristics of the datasets from each spectrograph. Table~\ref{tab:ObservationalSummary} summarises the key properties of each observation, including wavelength coverage, resolution, exposure time, velocity smearing induced by the exposure time, and transit baseline.

\begin{figure*}[htbp]
    \centering
    \includegraphics[width=\textwidth]{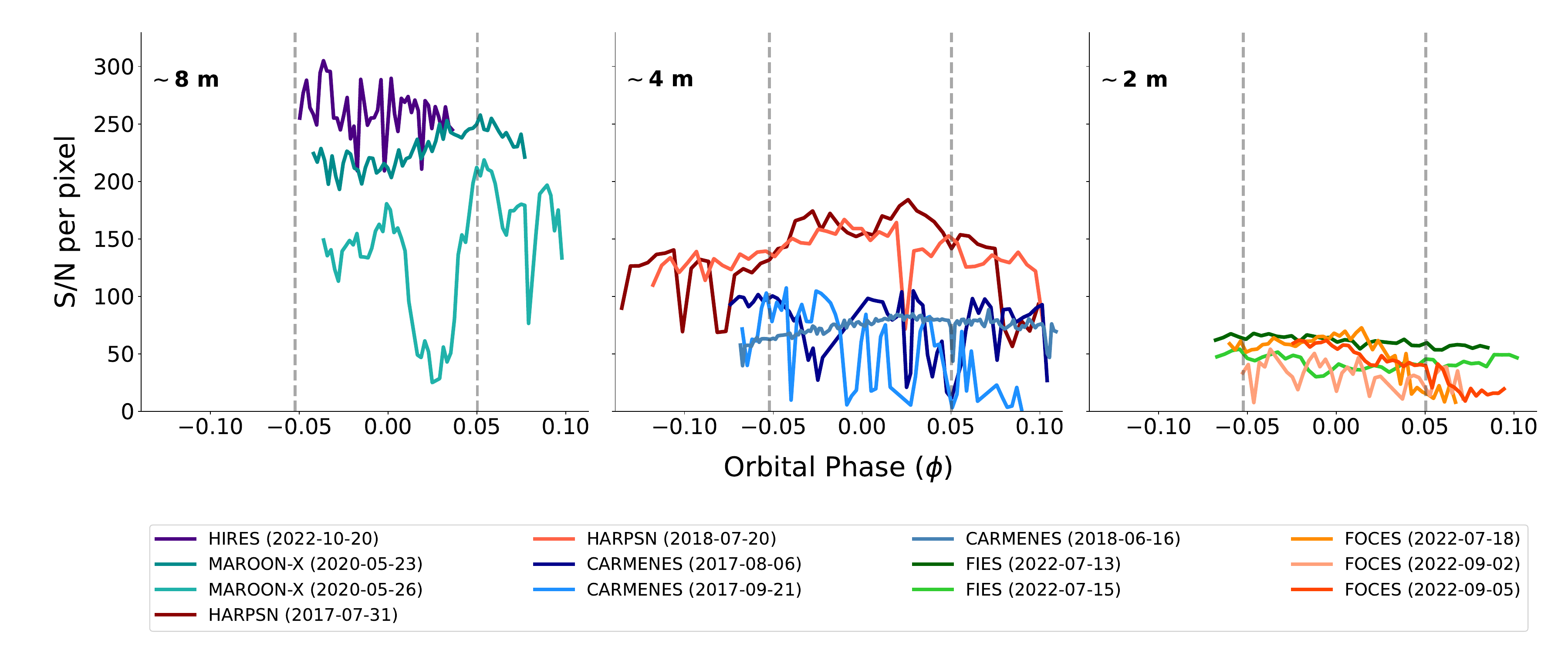}
    \caption{Phase-dependent signal-to-noise ratio (S/N) per pixel across all 13 datasets. The calculation is described in Sect.~\ref{sect:observations_and_data}. The datasets are grouped by approximate telescope aperture and shown on a shared S/N scale. Faded vertical dashed lines mark the start and end of transit.}
    \label{fig:true_snr_grid}
\end{figure*}

The phase-dependent data quality of the spectra is assessed in Fig.~\ref{fig:true_snr_grid}, where we compute a per-pixel S/N estimate for each observing night using the \texttt{DER\_SNR} algorithm~\citep{Stoehr_dersnr_2008} over the central 80\% of a representative echelle order near the centre of each spectrograph's wavelength range (typically $\approx 500$--$750$~nm). \texttt{DER\_SNR} calculates the S/N directly from the spectral flux without requiring continuum normalization or an explicit noise model. The algorithm assumes normally distributed noise uncorrelated at two-pixel separations and a locally linear continuum over short wavelength intervals. These measurements provide the observational S/N comparison between nights and trace changes due to observing conditions, airmass, and throughput during each transit.

\subsection{Dataset: TNG/HARPS-N}
Two transit observations used in this work come from archival observations of KELT-9b taken with the HARPS-N spectrograph on the 3.6\,m Telescopio Nazionale Galileo telescope. The spectrograph is a fibre-fed, cross-dispersed echelle spectrograph covering a wavelength range between 387 to 690\,nm at a spectral resolution of 115,\,000 \citep{Mayor_2003}. These observations used a second fibre for simultaneous wavelength calibration. We downloaded the pipeline-reduced spectra from the TNG archive. These data were originally taken on 31 July 2017 and 20 July 2018, and were first presented in \citep{Hoeijmakers_2018}, with subsequent papers utilising them for additional results \citep{Hoeijmakers_2019,Casasayas_Barris_2019,Yan_2019,Rainer_2019,Fisher_2020,Wyttenbach_2020,Asnodkar_2022a,Asnodkar_2022b,Borsato_2023,Borsato_2024,DArpa_2024,Stangret_2024}.

Both observations contain a large amount of baseline data, with a consistent exposure time of 600\,s for each exposure. For this exposure time, the Doppler smearing due to the planet’s orbital velocity is approximately 7.3\,km\,s$^{-1}$, corresponding to roughly three HARPS-N resolution elements (with $\Delta v \simeq c/R \approx 2.6\,\mathrm{km\,s^{-1}}$). Data quality for both nights remains relatively consistent across the transits, with the second night experiencing a temporary drop in signal strength during a small part of the transit. As a spectrograph, HARPS-N has good coverage over the visual spectrum, with strong throughput at optical wavelengths and excellent instrumental stability.

\subsection{Dataset: CAHA/CARMENES}
Three transit observations of KELT-9b using the CARMENES spectrograph on the 3.5\,m CAHA telescope are available from the Calar Alto Observatory archive. CARMENES is a fibre-fed, cross-dispersed echelle spectrograph, covering an optical arm between 520 to 960\,nm and an infrared arm between 960 to 1710\,nm \citep{Quirrenbach_2016_CARMENES}. For our analysis, we focus only on the visible channel (VIS). The three observations were taken on 6 August and 21 September 2017~\citep{Yan_2018,Yan_2019}, and 16 June 2018~\citep{Turner_2020}.

The CARMENES observations cover the entire transit, with the first two transits obtained using a consistent exposure time of 300\,s and the third using 111\,s exposures. For these exposure times, the Doppler smearing due to the planet’s orbital velocity is approximately 3.6\,km\,s$^{-1}$ and 1.35\,km\,s$^{-1}$, respectively, corresponding to roughly 1.15 and 0.43 resolution elements of CARMENES (with $\Delta v \simeq c/R \approx 3.17\,\mathrm{km\,s^{-1}}$). During these observations, simultaneous sky monitoring was implemented with a second fibre.

The first two nights have low signal-to-noise ratio (S/N) throughout the transit, while the third night achieves a signal-to-noise ratio comparable to the HARPS-N observations. During favourable observing conditions, CARMENES performs well over its wavelength range; however, its wavelength coverage begins in the red, extending into the near-infrared. This makes the spectrograph effective for resolving volatile species with strong transitions in this region, such as Ca\,\textsc{ii} and H\,\textsc{i}, but less sensitive to refractory metals, whose strongest lines lie at shorter wavelengths.

\subsection{Dataset: NOT/FIES}
We observed two additional transits of KELT-9b using the FIES spectrograph located on the Nordic Optical Telescope (NOT). The telescope and spectrograph have already been shown to be successful in cross-correlation studies like this \citep{Bello_Arufe_2022}. FIES is a fibre-fed, cross-dispersed high-resolution echelle spectrograph spanning a wavelength coverage of 368 to 898\,nm, with a mirror size of 2.56\,m \citep{Telting_2014_FIES}.

The FIES observations were obtained using a consistent exposure time of 500\,s, chosen to optimise photon noise given the smaller telescope aperture, resulting in an approximate signal-to-noise ratio of 50 per exposure. These choices were made to optimise the overall observing strategy \citep{Boldt_Christmas_2024}. For this exposure time, the Doppler smearing due to the planet’s orbital velocity is approximately 6.1\,km\,s$^{-1}$, corresponding to roughly 1.35 resolution elements of FIES (with $\Delta v \simeq c/R \approx 4.5\,\mathrm{km\,s^{-1}}$). These observations used FIES in its high-resolution, single-fibre configuration, with no simultaneous sky or reference channel.

Observations were taken on the 13th and 15th of July 2022. We opted to take the calibration frames specified in the FIES cookbook\footnote{\href{https://www.not.iac.es/observing/cookbook/current/Cookbook.php?instrument=FIES}{FIES Cookbook}}, and used FIEStool to reduce the data to its wavelength-calibrated two-dimensional echelle spectra \citep{Telting_2014_FIES}. For the first night of observations, conditions were ideal throughout the night. However, on the second night, thin cirrus clouds led to a reduction in signal quality through the night.

\subsection{Dataset: Wendelstein/FOCES}
Three additional nights originate from FOCES, a fibre-fed echelle spectrograph mounted on the 2.1\,m telescope at Wendelstein Observatory. The instrument spans a wavelength range of 383 to 885 nm at a resolution of R $\sim$ 70,000 \citep{Pfeiffer_1998, Brucalassi_et_2012, Wang_et_al_2017}. The wavelength-calibrated spectra were reduced using the instrument's standard pipeline, and results from the observations were first published by \cite{Borsato_2024}.

The Wendelstein observations were obtained using a consistent exposure time of 300\,s for each exposure. For this exposure time, the Doppler smearing due to the planet’s orbital velocity is approximately 3.6\,km\,s$^{-1}$, corresponding to roughly 0.85 resolution elements of FOCES (with $\Delta v \simeq c/R \approx 4.3\,\mathrm{km\,s^{-1}}$). These observations used a second fibre for simultaneous sky monitoring.

The three FOCES nights consistently contain the lowest S/N compared to the other telescopes. \cite{Borsato_2024} reported that while the lower signal is partially attributed to the smaller aperture of the telescope, consistent cloud coverage interfered with the majority of observations in this set. We keep these transits in our analysis for completeness, relying on our weighting methods (see Sect.~\ref{sect:weights}) to account for the overall data quality.

\subsection{Dataset: Keck/HIRES}\label{sect:keckhires}
The most recent transit in this study was observed using the HIRES spectrograph mounted on the 10\,m Keck telescope. HIRES is a slit-based, in-plane echelle spectrometer with a grating post-dispersion, providing a resolution of 67,\,000 and spanning a spectral range from 300\,nm to 1100\,nm. However, users must position the mosaic of 1 x 3 CCDs to select a specified wavelength region \citep{Vogt_1994_HIRES_Spectrograph}. The unique advantage of this spectrograph is that it allows choosing a wavelength region close to near-UV wavelengths. We used the HIRES echelle simulator to select a wavelength range from 316 to 600\,nm, aiming to go as blue as possible while still obtaining a sufficiently strong signal for cross-correlation.

The HIRES observations were obtained with a consistent exposure time of 200\,s, yielding an approximate signal-to-noise ratio of 75 per exposure given the telescope aperture. For this exposure time, the Doppler smearing due to the planet’s orbital velocity is approximately 2.4\,km\,s$^{-1}$, corresponding to roughly 0.4 resolution elements of HIRES (with $\Delta v \simeq c/R \approx 6.0\,\mathrm{km\,s^{-1}}$). This strategy follows previous recommendations while avoiding excessive losses to readout time \citep{Boldt_Christmas_2024}.

The reduction pipeline, MAKEE \citep{code_MAKEE}, designed for HIRES, automatically removes the blaze profile from the spectral orders when available. For cross-correlation analysis, preserving the blaze profile provides an approximate relative-throughput weighting across each spectral order, so that low-throughput order edges contribute less to the cross-correlation summation than high-throughput order centres. While this does not provide a complete noise model (e.g. omitting read-noise, dark current, and sky-background), it approximately weights the signal by the expected relative throughput. We therefore reintroduced the blaze profiles by using the measured master flat calibration frames, extracting each spectral order, computing a one-dimensional filter to capture the blaze profile, and applying this filter to the data.

\subsection{Dataset: Gemini-N/MAROON-X}
MAROON-X is a fibre-fed, high-efficiency echelle spectrograph designed for high-precision radial velocity measurements, tailored initially for detecting Earth-mass planets around mid-to-late M-dwarfs \citep{Seifahrt_2018_MAROONX, Seifahrt_2020_MAROONX}. The instrument is mounted on the 8\,m Gemini North telescope at Mauna Kea, Hawaii. It covers a wavelength range of 491 to 921\,nm at a resolution of $R \sim 85{,}\,000$.

Two transit observations were taken on 23 and 26 May 2020 as part of programme GN-2020A-Q-234 (PI: Bean). The data were reduced using the \texttt{MAROONXDRP} pipeline \citep{Seifahrt_2020_MAROONX}, which produces wavelength-calibrated echelle orders. These observations used a second fibre for simultaneous wavelength calibration. The MAROON-X observations were obtained using a consistent exposure time of 200\,s for both nights. For this exposure time, the Doppler smearing due to the planet’s orbital velocity is approximately 2.4\,km\,s$^{-1}$, corresponding to roughly 0.65 resolution elements of MAROON-X (with $\Delta v \simeq c/R \approx 3.7\,\mathrm{km\,s^{-1}}$).

In terms of data quality, the 23 May dataset exhibits the highest S/N of all nights in our sample, with good stability throughout the transit. In contrast, the 26 May dataset has a lower S/N overall and exhibits a pronounced flux drop midway through the night due to intermittent cloud cover.

\begin{table*}
    \centering
    \caption{Summary information of the transit observations used in this study}
    \label{tab:ObservationalSummary}
    \scriptsize
    \setlength{\tabcolsep}{3pt}
    \begin{tabular*}{\textwidth}{@{\extracolsep{\fill}} l l l c c l r c c c @{}}
    \hline \hline
        Date [dd-mm-yyyy] & Spectrograph & Feed type & Dim ["] & Tel. [m] & Range [nm] & Resolution & Exp. [s] & N & Base. [\%] \\ \hline

                20-07-2018 & HARPS-N & Fibre & 1.0 & 3.58 & 387--690 & 115\,000 & 600 & 46 & 50\\
        31-07-2017 & HARPS-N & Fibre & 1.0 & 3.58 & 387--690 & 115\,000 & 600 & 49 & 57\\

        16-06-2018 & CARMENES & Fibre & 1.5 & 3.5 & 520--960 & 94\,600 & 111 & 50 & 39\\
        21-09-2017 & CARMENES & Fibre & 1.5 & 3.5 & 520--960 & 94\,600 & 300 & 50 & 28\\
        06-08-2017 & CARMENES & Fibre & 1.5 & 3.5 & 520--960 & 94\,600 & 300 & 54 & 39\\

        15-07-2022 & FIES & Fibre & 1.5 & 2.56 & 368--898 & 67\,000 & 500 & 39 & 36\\
        13-07-2022 & FIES & Fibre & 1.5 & 2.56 & 368--898 & 67\,000 & 500 & 36 & 30\\

        05-09-2022 & FOCES & Fibre & 1.5 & 2.1 & 380--885 & 70\,000 & 300 & 39 & 17\\
        02-09-2022 & FOCES & Fibre & 1.5 & 2.1 & 380--885 & 70\,000 & 300 & 41 & 33\\
        18-07-2022 & FOCES & Fibre & 1.5 & 2.1 & 380--885 & 70\,000 & 300 & 41 & 18\\

        20-10-2022 & HIRES & Slit & 0.574 & 10 & 316--600 & 67\,000 & 200 & 46 & 2\\

        26-05-2020 & MAROON-X & Fibre & 0.77 & 8.1 & 491--921 & 85\,000 & 200 & 62 & 29\\
        23-05-2020 & MAROON-X & Fibre & 0.77 & 8.1 & 491--921 & 85\,000 & 200 & 58 & 16\\
    \hline
    \end{tabular*}
    \tablefoot{This table summarises the broad characteristics of each observing site and observation. Columns show the date, spectrograph name, feed type (slit vs. fibre), on-sky dimension (slit width or fibre diameter), telescope aperture size, wavelength range, resolution, exposure time, number of exposures, and total out-of-transit baseline as a percentage.}
\end{table*}

\section{Methods}\label{sec:methods}
This section describes the modelling, signal-extraction, and statistical frameworks adopted in this work. We first outline how we measure the radial velocity (\vsys) and rotational velocity (\vsini), and how the spectra are prepared for cross-correlation. We then describe the cross-correlation framework used to extract planetary signals with standardised templates, including the construction of velocity–velocity diagrams and the removal of residual stellar signals. We then turn to the treatment of residual structures in the CCFs, describe how forward models are injected into the data, and how each observation is weighted and combined.

Finally, we demonstrate the issue of velocity-axis correlation when applying the cross-correlation technique, outline how it arises in velocity-gridded analyses, and present a solution. We conclude by presenting our bootstrap-based approach, which addresses velocity-axis correlation and yields more reliable estimates of the forward-model parameters and their associated uncertainties.

\subsection{Systemic and rotational velocities of KELT-9A}
\label{sect:vsys}
In cross-correlation studies of exoplanet atmospheres, it is common to retrieve a radial-velocity offset that is sometimes interpreted as the systemic velocity. In practice, these offsets are often displaced from the stellar $v_{\mathrm{sys}}$, and this discrepancy has been linked to species-dependent atmospheric dynamics, pressure levels, and orbital phase~\citep{Merritt_2021,Kesseli_2022,Seidel_2025,Prinoth_2025}. For this reason, we independently measure the stellar $v_{\mathrm{sys}}$ from the photospheric lines, allowing a direct comparison between the stellar value and the velocity offsets recovered from our atmospheric detections.

Measuring the systemic velocity of KELT-9A is challenging due to its rapid rotation. The resulting strong rotational broadening makes it difficult to determine line centres using simple Gaussian fits. As a result, we use least-squares deconvolution (LSD), an established method for constructing high-S/N mean line profiles from many spectral lines~\citep{Donati_1997,Kochukhov_2010}. LSD assumes that the selected spectral lines share a common intrinsic shape, while allowing their depths and central wavelengths to vary. The method then determines the statistical weighting that best reproduces the observed spectrum. The resulting composite profile captures the average behaviour of the stellar absorption lines and is commonly adopted for radial-velocity measurements of rapidly rotating stars~\citep{Borsa_2019,Asnodkar_2022b}. In this work, we perform the LSD analysis using the \texttt{SpecpolFlow} software package~\citep{Folsom_SpecpolFlow_2025}.

For each observation, we extracted and analysed the out-of-transit exposures when available. We processed each spectrum using LSD. To prevent continuum-merging systematics from introducing systematic velocity shifts across the multi-instrument dataset, the echelle data were continuum-normalised on an order-by-order basis. Independent normalisation was applied to each raw order using a flexible polynomial model. Furthermore, order edges were trimmed to prevent detector-boundary artefacts from projecting into the final Least Squares Deconvolution (LSD) profile. The LSD profile was computed for each order independently, and these profiles were then combined using their inverse-variance weights to construct the final composite profile for each exposure. A line mask was generated using the Vienna Atomic Line Database (VALD; \citep{Ryabchikova_2015}), based on stellar parameters similar to KELT-9A ($T_{\mathrm{eff}} = 10{,}000$K, $\log g = 4.0$, [Fe/H] = 0.0,~\citep{Gaudi_2017}). Heavily saturated lines such as the Balmer series, the Ca\,\textsc{ii}\,K line, and (where applicable) the Ca\,\textsc{ii} near-infrared triplet were excluded from the mask. We used the \texttt{cleanMaskUI} module of \texttt{SpecpolFlow} to define broad wavelength exclusion intervals around these strong features (typically spanning $\sim 500{-}1000\mathrm{\,km\,s^{-1}}$). This module removes all template lines falling within these broad windows from the final line mask, ensuring that no weak template lines are included near the damping wings of strong features during the LSD inversion and preventing weak lines from fitting residual strong-line wings. The stellar template was refined to ensure model purity by removing specific transitions identified as non-photospheric. Specifically, the Sodium doublet was removed from the mask across all observations to remediate strong Interstellar Medium (ISM) absorption that was found to introduce contaminating substructure. We further audited the resulting LSD profiles and residuals, particularly for the high-S/N MAROON-X and HARPS-N datasets, to check for potential mismatches between the VALD linelist and the observations. No coherent residual structures were identified that would indicate significant errors or omissions in the stellar template. LSD profiles were then computed on velocity grids spanning $-250$ to $+250\mathrm{\,km\,s^{-1}}$. The per-pixel flux errors supplied by the reduction pipelines were utilised as inverse-variance weights within the LSD solver, ensuring that low-quality data regions---including core-saturated lines and telluric-contaminated pixels---are naturally marginalised based on their local statistical significance. This framework aims to accurately propagate uncertainties to account for statistical noise and instrumental stability.

Each LSD profile was fit with a forward model based on the analytic rotational broadening formalism of \cite{Gray_2005}. The model describes the stellar LSD mean profile assuming a delta-function (zero-width) intrinsic line profile rather than a Gaussian or Voigt profile. We implement this model using the \texttt{RotBroadProfile} class from the \texttt{PyAstronomy} library \citep{pya}. The model is parameterised by the amplitude ($A$), projected rotational velocity (\vsini), linear limb-darkening coefficient ($\varepsilon$), and line-centre velocity ($\mu$).

\begin{equation}
    G(v) = \frac{2(1-\varepsilon)\sqrt{1-\left(\frac{v}{v_{\mathrm{rot}}\sin i}\right)^2} + \frac{\pi\varepsilon}{2}\left[1-\left(\frac{v}{v_{\mathrm{rot}}\sin i}\right)^2\right]}{\pi\,(v_{\mathrm{rot}}\sin i)\,(1-\varepsilon/3)} ,
\end{equation}

where $v_{\mathrm{rot}}\sin i$ sets the half-width of the rotational profile. The model was fit as an empirical forward model to each stellar LSD profile. Uniform priors were adopted, bounded within the ranges ($A \in [-3,0]$, $v_{\mathrm{rot}}\sin i \in [80,140]$~km\,s$^{-1}$, $\varepsilon \in [0,1]$, $\mu \in [-35,-15]$~km\,s$^{-1}$), where $A$ is the integrated area and the lower bound of $-3$ corresponds to a peak line depth of $\approx 1.7\%$. Predictive uncertainties for the modelled LSD profile of each stellar exposure were obtained by propagating the fitted-parameter uncertainties through the same forward model.

To combine observations across all nights and instruments, we sampled each radial velocity measurement 1000 times within its uncertainty before combining all measurements into a single distribution. When combining across all nights, this aggregation was performed as an inverse-variance weighted bootstrap resampling to account for the varying data quality between different nights. The mean of the resulting distribution was then calculated, and its uncertainty estimated using $10^{4}$ bootstrap iterations to obtain the final measurements of \vsys \,and \vsini.

\subsection{Preparing the spectra for cross-correlation of KELT-9b's atmosphere}
\label{sect:preparing_spectra}
Preparation of the data for cross-correlation was implemented using the \texttt{tayph} software package~\citep{Hoeijmakers_tayph}. To prepare and process the data for cross-correlation, we first shifted each spectrum to the Earth rest frame. This was achieved by measuring the bulk instrumental velocity drift by cross-correlating telluric features and applying the resulting scalar velocity offset as a global Doppler shift to all echelle orders simultaneously. We then used our one-dimensional spectra obtained from Sect.~\ref{sect:vsys} to create a master spectrum for each night of observation. From each master spectrum, we identified clean regions that contained telluric features with well-defined continua. Using these regions, we fit the tellurics in each exposure utilising the software package \texttt{Molecfit}~\citep{Smette_2015,Kausch_2015}, which solves the radiative transfer equations for Earth’s atmosphere, and then removed the tellurics by dividing out the fitted spectrum from the data. The corresponding wavelength intervals are listed in Table~\ref{tab:telluric_regions}. We applied this division to both the pre-normalised two-dimensional spectra and the one-dimensional continuum-normalised spectra. Applying the correction to the pre-normalised spectra is valid because telluric removal and blaze normalisation are multiplicative processes.

After removing tellurics, we corrected for the Earth’s barycentric motion by computing the expected radial velocity shift at each observing time and location. We also corrected for the star's reflex Keplerian motion induced by the planet using its known radial velocity semi-amplitude of 293.0\,m\,s$^{-1}$~\citep{Borsa_2019}. Finally, we removed the stellar component by computing a time-averaged spectrum from the out-of-transit spectra on an order-by-order basis, and normalising each exposure by this average.

Differences in continuum level between exposures can produce non-flat residuals after this step. To correct for this, we apply a colour correction using a high-pass filter with a width of 80 km\,s$^{-1}$ in velocity space, which maps the broadband continuum shape and flattens it. This filtering is applied after removal of the time-averaged stellar component and is intended to remove low-frequency residual normalisation structure. The adopted window is deliberately broader than the planetary absorption profile, allowing broad residual fluctuations to be removed while preserving the narrower atmospheric CCF signal. Finally, for the normalised residuals, we apply a running median absolute deviation filter with a width of 20 pixels to flag spurious data values arising from detector artefacts. All flux values more than 5$\sigma$ above the local median are flagged as bad pixels and replaced via linear interpolation. Because the flagged regions are short and isolated, linear interpolation preserves the local continuum without affecting the cross-correlation signal. We then visually inspect each order for any remaining contaminants, such as interstellar absorption or residuals from deep telluric lines, and mask the affected wavelength regions with NaNs. During our visual inspection, the only interstellar medium (ISM) features consistently prominent enough across all instruments to require manual masking were the narrow absorption spikes of the Sodium (Na\,I) doublet; these are summarised in Table~\ref{tab:ism_features} in Appendix~D. The fraction of flux values masked in this procedure is summarised in Table~\ref{tab:badpix_stats}. The largest masked fractions occur for CARMENES and HIRES. These high values are driven by broad wavelength/order exclusions rather than isolated pixel clipping. Detailed descriptions of the specific telluric and instrumental artefacts requiring these conservative masks, along with representative diagnostic figures (Figs.~\ref{fig:carmenes_mask_diagnostic} and \ref{fig:hires_mask_diagnostic}), are provided in Appendix~\ref{app:masked_pixels}. This masking sacrifices some flux information, but prevents contaminated regions from projecting coherently into the CCF and producing artificial atmospheric signals.

This procedure is performed automatically within the \texttt{tayph} framework, and has been successfully applied in other transmission spectroscopy studies of ultra-hot Jupiter atmospheres (e.g. \citealt{Hoeijmakers_2020,Damasceno_2024,Seidel_2025}). We illustrate the effects of the cleaning processes in cross-correlation space, where the progressive removal of systematics and the recovery of the planetary signal is directly visualised in Fig.~\ref{fig:ccf_cleaning}.

\subsection{Cross-correlation framework}
To search for atomic and ionised species within the atmosphere of KELT-9b, we employ the cross-correlation technique \citep{Snellen_2010_CC}. Cross-correlation is a series of weighted averages of the expected line positions of a species that may exist in the atmosphere of an exoplanet, which are shifted over a grid of Doppler velocities. As the planet's radial velocity shifts during the transit, the weighted average is maximised when the line positions match the planet's radial velocity at the time the observation was taken. Mathematically, it is defined as;

\begin{equation}
    c(v) = \sum_{i=0}^{N} x_i \hat{T}_i(v)
\label{eq:ccf}
\end{equation}

where $c(v)$ is the weighted average of the flux values, $x_i$ refers to the flux values of each data point, and $\hat{T}_i(v)$ is a template spectrum containing the line positions to be converted to weights for the summation from the following formula. In Eq.~\ref{eq:ccf}, the flux values $x_i$ retain the relative blaze-profile shape, so that low-throughput pixels contribute less than high-throughput pixels to the cross-correlation summation.

\begin{equation}
    \hat{T}_i(v) = \frac{T_i(v)}{\Sigma T_i(v)}
\end{equation}

where $\hat{T}_i(v)$ is the normalised template spectrum, $T_i(v)$ represents the original template spectrum values, and $\sum T_i(v)$ is the sum of all values in the original template spectrum. The template spectra used for cross-correlation are typically generated using standard chemistry packages such as \texttt{FASTCHEM}~\citep{Stock_2018,Stock_2022,Kitzmann_2024}, which can be combined with radiative transfer packages such as~petitRADTRANS~\citep{Molliere_2019} or \texttt{PYRAT BAY}~\citep{Cubillos_2021} to produce atmospheric templates for cross-correlation. One issue with developing one’s own templates is that models vary across studies, which can complicate reproducibility. The MANTIS templates~\citep{Kitzmann_2021} are a library of cross-correlation templates that cover many atomic and ionised species, created to standardise the templates used for species detection. Here we use these templates for our cross-correlation analysis.

A good initial guess for an appropriate temperature is to use the equilibrium temperature of the planet, for KELT-9b, this is $\sim$4000~K~\citep{Borsa_2019}, which has been shown to align well with detections of atomic iron in the planet’s atmosphere~\citep{Hoeijmakers_2018}. However, a recent detection of neutral oxygen in the planet’s atmosphere~\citep{Borsa_2022b_OIdetection} was obtained using a higher-temperature template. This can likely be explained by the fact that the dayside of KELT-9b is expected to reach temperatures significantly higher than the equilibrium temperature. In contrast, the nightside is expected to be cooler~\citep{Wong_2020}. The best-fit temperature likely varies on a species-by-species basis, due to the fact that different species probe different regions of the planet’s atmosphere, as well as limitations in the models used. For consistency with~\cite{Borsa_2022b_OIdetection}, we select templates generated at 5000\,K.

\subsubsection{Constructing velocity–velocity diagrams}
Velocity–velocity diagrams --- often referred to as $K_p$–\vsys maps (where $K_p$ is the planet’s radial velocity semi-amplitude) --- enhance the detection fidelity of the cross-correlation function by co-adding exposures into a single combined signal~\citep{Brogi_2012}. Assuming a static, circular orbit, the planet’s radial velocity can be expressed as:

\begin{equation}
v_r(\phi) = V_{\text{sys}} + K_p \sin(i) \sin(2\pi \phi),
\label{eq:rv_eq}
\end{equation}

where $v_r(\phi)$ is the radial velocity at orbital phase $\phi$, $V_{\text{sys}}$ is the systemic velocity, $K_p$ is the semi-amplitude of the planet’s radial velocity, and $i$ is the orbital inclination. The $K_p$–\vsys diagram is constructed by shifting the radial-velocity points of the CCF onto a grid of potential planetary rest frames and averaging over all transit exposures, yielding a two-dimensional array representing the cross-correlation strength as a function of $K_p$ and \vsys. For this work, we choose [0, 400] km\,s$^{-1}$ in steps of 1 km\,s$^{-1}$. The planetary signal appears as a localised peak where the assumed velocity matches the actual orbital motion. As ultra-hot Jupiters often exhibit inflated and non-static atmospheres, the apparent \vsys and $K_p$ are known to deviate from their nominal values~\citep{Ehrenreich_2020,Prinoth_2022}.

\subsection{Removal of the Doppler shadow}
\label{sect:dopplershadow}
When a CCF is produced with a template containing photospheric species, a bright time-dependent feature caused by the Rossiter--McLaughlin (RM) effect is often present. This feature, commonly called the Doppler shadow, must be removed before measuring the planetary atmospheric signal. To model the Doppler shadow, we use \texttt{StarRotator}~\citep{Hoeijmakers_StarRotator} to generate exposure-resolved models for each species, spectrograph bandpass, and observing night (yielding 377 individual models across our 29 detected species and 13 datasets). The Doppler-shadow trajectory is fixed by the known spin--orbit geometry, leaving only the overall scaling factor to be fitted to the CCF before subtraction. The phase interval where the planetary and RM signals overlap ($-0.015 \le \phi \le 0.001$) is excluded from both the Doppler-shadow fit and the planetary analysis. Those exposures are assigned zero weight during combination. Figure~\ref{fig:ccf_cleaning} illustrates the Doppler-shadow model and its subtraction.

\subsection{Detrending residual structures in CCFs}
After removing the Doppler shadow, residual signals may still contaminate the cross-correlation function. These have been attributed to sources such as imperfect telluric correction \citep{Prinoth_2022,Borsato_2023,Borsato_2024}, stellar pulsations \citep{Hoeijmakers_2019,Wyttenbach_2020}, and dynamic atmospheric variability or flare-driven mass-loss events \citep{Cauley_2019}. Such residuals introduce spurious features into velocity--velocity diagrams, complicating signal interpretation and interfering with the fitting procedures described in Sect.~\ref{sect:fitting_the_ccf}. To isolate the planet's signal and enhance model performance, we apply an additional cleaning step.

It is possible to remove residual structures from the CCF using vertical detrending approaches~\citep{Prinoth_2022,Prinoth_2023,Borsato_2025}. Using the radial velocity equation (Eq.~\ref{eq:rv_eq}), we trace the expected path of the planetary signal across the cross-correlation map. We then mask a 20\,km\,s$^{-1}$ window on either side of this path, as shown in Panel~H of Fig.~\ref{fig:ccf_cleaning}, to exclude the planet's trail from the fit. For each velocity step in the CCF, we fit a fifth-order polynomial along the orbital phase direction to the remaining (unmasked) pixels. We determined the optimal polynomial degree by evaluating the trade-off between background noise removal and injected signal preservation (see Appendix~\ref{app:poly_order_diagnostics}). While the background RMS decreases rapidly up to degree~5, higher degrees offer diminishing returns in noise reduction while progressively distorting the underlying planetary signal. We therefore adopted a fifth-order polynomial for all vertical detrending. This produces a two-dimensional model of the residual structure, shown in Panel~I. Although polynomial detrending is performed independently at each radial-velocity step, the resulting trends vary smoothly between neighbouring velocities, thereby maintaining correlations along the radial-velocity direction. Finally, we subtract this model from the Doppler-shadow-removed CCF to produce a cleaned version, shown in Panel~J of Fig.~\ref{fig:ccf_cleaning}.

\subsection{Model injection}
To assess whether a non-detection arises from an absence of signal or from analysis limitations, we employ model injection—a standard technique in high-resolution spectroscopy \citep{Snellen_2010_CC,Birkby_2013,Hoeijmakers_2018_tau_bootis}. A forward model of the planet’s atmosphere is rotationally and instrumentally broadened, interpolated to the spectrograph wavelength grid, and added to the pipeline-reduced spectra. Cross-correlating this injected data yields a combined data + model signal. Subtracting this from the original CCF isolates the processed-model contribution, allowing us to test whether a real signal would be recoverable under the applied analysis steps. As an additional control, we injected the Fe\,I model at negative $K_p$ and passed it through the same cleaning, detrending, and weighted-combination steps as the science CCFs. The injected signal is recovered on the negative-$K_p$ side of the combined velocity--velocity diagram (Fig.~\ref{fig:negative_kp_injection_control}), showing that the cleaning procedure does not erase a planetary-like signal or force it to the expected positive-$K_p$ solution.

\subsection{Mitigating velocity-axis correlation in velocity-gridded fits}
\label{app:velocity_axis_correlation_bootstrap}
Cross-correlation functions (CCFs) and least-squares deconvolution (LSD) profiles are computed via a grid search over discrete radial velocity steps. Owing to the finite spectral resolution of the instrument, the changing radial velocity of the planet during transit, and the interpolation required to shift spectra and templates between reference frames, neighbouring velocity bins in the resulting CCF may not be statistically independent. This induces correlations between adjacent velocity steps, a situation referred to here as velocity-axis correlation. In the presence of velocity-axis correlation, fitting statistics derived from forward models can yield biased uncertainty estimates and inflated detection significances.

To demonstrate this effect, and to motivate our proposed mitigation strategy, we introduce a simple toy model. Consider a simple Gaussian line combined with noise.

\begin{equation}
    y_i = A \exp\!\left[-\frac{(v_i - \mu)^2}{2\sigma^2}\right] + \epsilon_i,
\end{equation}

where $A,\,\mu$, and $\sigma$ are the amplitude, line centre, and line width, $\epsilon_i \sim \mathcal{N}(0, \sigma_{\mathrm{noise}}^2)$, and the velocities are uniformly spaced as $v_i = i\,\Delta v$. If we define a forward model of this Gaussian function, we can estimate the amplitude with an associated uncertainty. For any fitting routine, there exists a minimum uncertainty that can be achieved, which is set by $\epsilon_i$. We can derive this uncertainty using the Fisher formalism to calculate the sensitivity of the data.

\begin{equation}
    \mathcal{I}(A)
    = \frac{1}{\sigma_{\mathrm{noise}}^2}
      \sum_i \left(\frac{\partial f_i}{\partial A}\right)^2
    = \frac{1}{\sigma_{\mathrm{noise}}^2}
      \sum_i e^{-\frac{(v_i - \mu)^2}{\sigma^2}}.
\end{equation}

\noindent
Approximating the sum as a Riemann integral yields

\begin{equation}
    \sum_i e^{-\frac{(v_i - \mu)^2}{\sigma^2}}
    \approx \frac{1}{\Delta v}\int_{-\infty}^{\infty}
    e^{-\frac{(v - \mu)^2}{\sigma^2}}\,\mathrm{d}v
    = \frac{\sqrt{\pi}\,\sigma}{\Delta v},
\end{equation}

\noindent
leading to the Cramér--Rao bound~\citep{Rao_1945,Cramer_1946}, which gives a lower bound on the variance, $\mathrm{Var}(A)\ge \mathcal{I}(A)^{-1}$, and hence the minimum achievable uncertainty on $A$,

\begin{equation}
    \sigma_A =
    \sigma_{\mathrm{noise}}
    \sqrt{\frac{\Delta v}{\sqrt{\pi}\,\sigma}}.
\end{equation}

This expression represents the theoretical lower limit for $\sigma_A$ under the assumption of statistically independent samples, predicting that uncertainty scales as $\sigma_A \propto \sqrt{\Delta v}$. In practice, however, velocity bins below the instrument’s native sampling interval $\Delta v_0$ are not independent. To account for this, we define an effective sampling step,

\begin{equation}
    \Delta v_{\mathrm{eff}} = \max(\Delta v,\, \Delta v_0),
\end{equation}

\noindent
which limits the Fisher information to the true information content of the data. The corresponding analytic uncertainty relation becomes

\begin{equation}
    \sigma_A(\Delta v) =
    \sigma_{\mathrm{noise}}
    \sqrt{\frac{\max(\Delta v, \Delta v_0)}{\sqrt{\pi}\,\sigma}},
\end{equation}

\noindent
predicting a floor at fine sampling ($\Delta v \leq \Delta v_0$) and a $\sqrt{\Delta v}$ rise when the data is undersampled.

To demonstrate the impact of velocity-axis correlation, we performed a numerical experiment. We generated a Gaussian line with $A=0.5$, $\mu=0$, and $\sigma=20~\mathrm{km\,s^{-1}}$, evaluated on a grid with $\Delta v_{\mathrm{true}}=4~\mathrm{km\,s^{-1}}$ over $v \in [-400,400]~\mathrm{km\,s^{-1}}$, and added Gaussian noise with a root-mean-square of 0.05 in normalised-flux units, corresponding to a Cramér–Rao amplitude minimum of $\sigma_A = 0.017$. These noisy data were then interpolated and resampled onto a series of target grids with spacings $\Delta v \in \{0.1,0.5,1,2,3,4,5,\ldots,30\}~\mathrm{km\,s^{-1}}$. Each resampled dataset was fit with a forward Gaussian model using the same JAX--NumPyro approach we have implemented, retrieving the posterior mean and standard deviation of $A$. We refer to this as the direct-fit approach.

As an alternative, $A$ can be estimated via bootstrapping. For each $\Delta v$, we jittered the data within their uncertainties before interpolating to the new grid. The model was then fit, and the best-fitting value of $A$ was stored. This process was repeated 1000 times to build a distribution of best-fit $A$ values for each grid spacing, from which the mean and standard deviation can be extracted as alternative estimates of the amplitude and its uncertainty. We term this the bootstrap-fit.

\begin{figure}[htbp]
    \centering
    \includegraphics[width=\columnwidth]{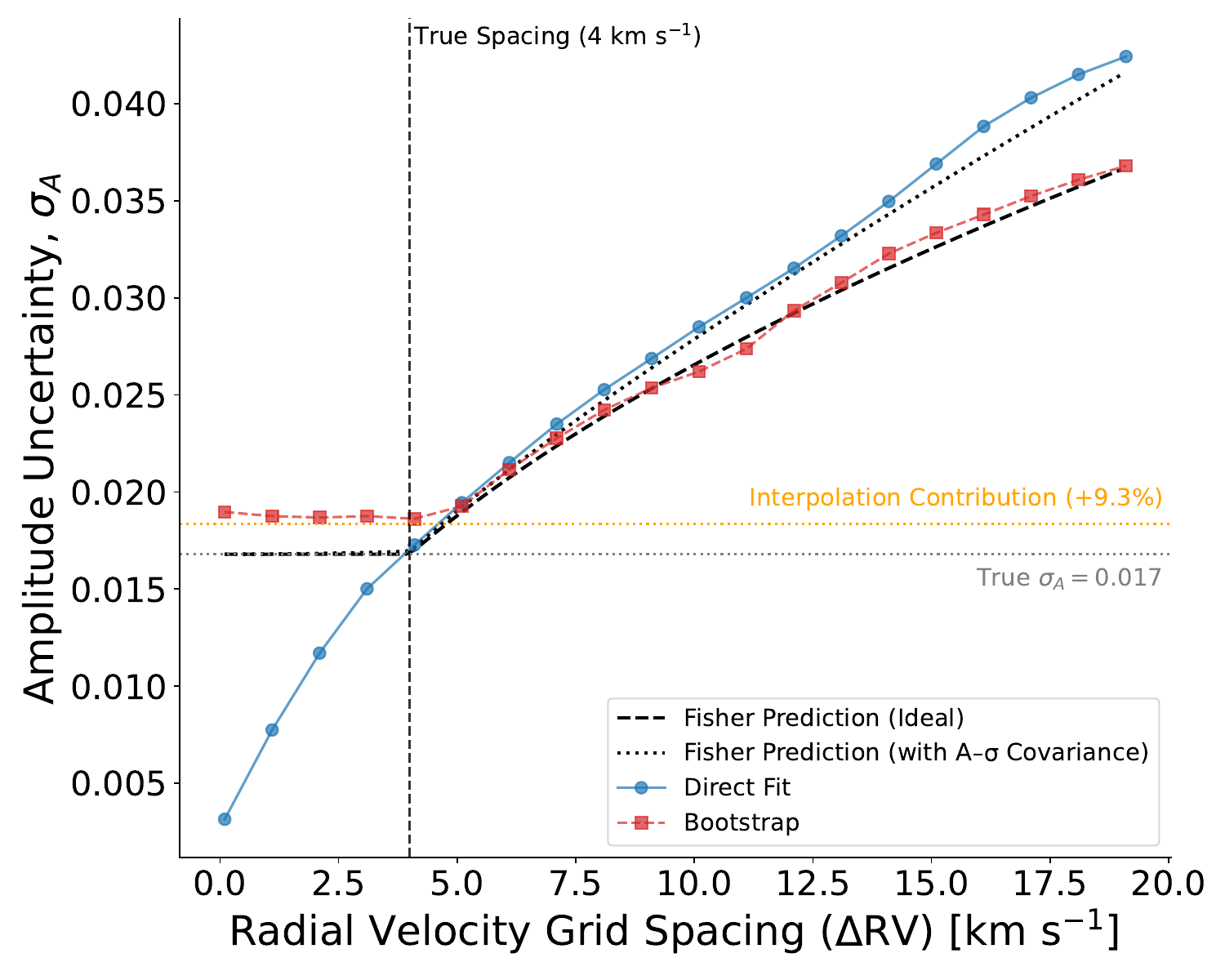}
    \caption{The amplitude uncertainty versus velocity-grid spacing from the two different fitting methods. Blue circles show the direct-fit amplitude uncertainty, while red squares show the bootstrap-fit uncertainty. The grey dash-dotted line indicates the analytic Fisher-information model with sampling saturation, while the orange dotted line marks the predicted inflation ($+9.3\%$) due to interpolation-induced correlations in the residuals. The purple curve includes an additional inflation term arising from the amplitude--width degeneracy, which becomes increasingly important at larger radial-velocity spacings. The vertical dashed line marks the native resolution of the simulated data ($\Delta v_{\mathrm{true}}=4~\mathrm{km\,s^{-1}}$). Together, these results demonstrate that bootstrap-derived uncertainties remain conservative and closely follow the combined analytic model, capturing both the loss of information from correlated sampling and the geometric degeneracy between amplitude and width in coarse-gridded fits.}
    \label{fig:amp_unc_vs_step}
\end{figure}

If we plot the uncertainty on $A$ ($\sigma_A$) as a function of the velocity grid spacing ($\Delta v$), as shown in Fig.~\ref{fig:amp_unc_vs_step}, the direct-fit $\sigma_A$ values (blue) follow the expected analytic $\sqrt{\Delta v}$ scaling. However, when the fit is performed on an interpolated grid finer than the actual sampling grid, the estimated uncertainty continues to decrease, even below the Cramér–Rao bound. This occurs because, in a formal $\chi^2$ fit, the precision scales with the number of data points. Interpolating onto an increasingly fine grid, therefore, artificially increases the apparent certainty of the measured value, since the underlying assumption of independent grid points is violated once the spacing drops below the intrinsic resolution. Crucially, this effect is continuous, with no noticeable change in the behaviour of the uncertainties, and therefore cannot be readily identified from the fit alone.

The bootstrap fit (red) consistently produces uncertainties larger than the Cramér–Rao bound and remains effectively constant when the sampling is finer than the actual grid spacing. It converges to the Fisher scaling once the grid spacing begins to undersample the data, beyond $\Delta v_{\mathrm{true}}=4~\mathrm{km\,s^{-1}}$. The larger uncertainty observed in the oversampled regime arises because, while the bootstrap jittering correctly captures the measurement noise, the interpolation step redistributes this noise across neighbouring samples, introducing weak correlations that reduce the effective number of independent points and consequently inflate the inferred uncertainty. This effect can be quantified using the lag-1 correlation coefficient of the residuals: $\rho_1$, which we derive in Appendix~\ref{app:interpolation_Noise}, measured after removing the smooth trend from the interpolated data. We find $\rho_1 \approx 0.1$, corresponding to a correlation-induced uncertainty inflation of roughly 9\%.

An additional difference between the two approaches emerges when the velocity grid becomes undersampled. In the direct-fit case, the amplitude becomes increasingly degenerate with the line width, and the loss of information under undersampling exacerbates this degeneracy, leading to a less precise estimate of the amplitude uncertainty. We derive an estimate of the degeneracy loss in Appendix~\ref{app:amplitude_degeneracy}. The bootstrap approach, by contrast, does not rely on the local covariance structure of the fitted model and is therefore less sensitive to this degeneracy, allowing the inferred uncertainty to remain close to the Fisher limit.

The results of this toy example demonstrate the limitations of performing direct fits to the data when $\Delta v$ is difficult to quantify, a common issue in high-resolution cross-correlation studies. Uncertainties risk becoming biased if the velocity grid is oversampled, while undersampling leads to a loss of confidence in the fitted parameters and results in uncertainties that are worsened by parameter degeneracies. Bootstrapping, on the other hand, can be performed on a fine radial velocity grid, which incurs only a slight increase in inferred uncertainty; however, these uncertainties are naturally integrated into the process. This indicates that the uncertainties and detection significances produced by this method are conservative, which is beneficial when evaluating the robustness of claimed detections.

\subsection{Modelling the cross-correlation function in a bootstrap framework}\label{sect:fitting_the_ccf}
Standard approaches to quantifying cross-correlation detections rely on z-statistics, evaluating the significance of the CCF peak in a given K$_p$–\vsys row. However, as shown in \citet{Borsato_2023}, significant peaks often appear at multiple locations due to the highly correlated structure of the velocity–velocity map. This velocity-axis correlation in both $K_p$ and \vsys directions artificially reduces uncertainties, thereby inflating detection significances.

Fitting a Gaussian to a single row of the K$_p$–\vsys diagram, as done in many previous works \citep[e.g.][]{Hoeijmakers_2019}, is sensitive to this correlation structure. While bootstrapping methods attempt to mitigate noise effects by sampling the background distribution, they remain vulnerable to degeneracies in selecting the “correct” row to fit.

Given these limitations, we use the K$_p$–\vsys diagram only as a visual check. Instead, we model the time-resolved cross-correlation function  directly, similar to the approach of \citet{Prinoth_2023,Hoeijmakers_2024}. This enables us to constrain the atmospheric signal without relying on assumptions about which row of the velocity–velocity diagram is optimal.

To weight the atmospheric absorption signal across orbital phases, we implement a linear limb-darkened Mandel--Agol transit light-curve profile \citep{Mandel_Agol_2002} computed using the \texttt{batman} package \citep{Kreidberg_2015} as our transit weighting function $W(\phi)$. The profile is computed using fixed literature orbital parameters ($a/R_\star = 3.153$, $i = 86.38^\circ$, $R_p/R_\star = 0.08228$) and a linear limb-darkening coefficient ($\varepsilon = 0.367583$). This provides a physical transit profile without introducing additional free parameters into the fit.

Our forward model describes the CCF trace using a set of seven free parameters: the absorption amplitude and its phase dependence ($A_0$, $A_1$), the systemic velocity (\vsys), the planetary orbital velocity semi-amplitude and its phase dependence ($K_{p,0}$, $K_{p,1}$), and the Gaussian width of the CCF trace and its phase dependence ($\sigma_0$, $\sigma_1$). The model is given by

\begin{equation}
    \mathrm{CCF}(v_r, \phi) =
    \frac{A(\phi)\,W(\phi)}{\sigma(\phi)\sqrt{2\pi}}
    \exp\left[
        -\frac{1}{2}
        \left(
            \frac{v_r - v_{\mathrm{model}}(\phi)}
                 {\sigma(\phi)}
        \right)^2
    \right],
\end{equation}

where the velocity of the planetary signal is

\begin{equation}
    v_{\mathrm{model}}(\phi)
    = V_{\mathrm{sys}} + K_p(\phi) \sin(2\pi\phi)\sin(i),
\end{equation}
\newline
\noindent
with $i = 86.38^\circ$, and the phase-dependent amplitude, width, and orbital velocity are

\begin{equation}
    A(\phi) = A_0 + A_1 \sin(2\pi\phi), \qquad
    \sigma(\phi) = \sigma_0 + \sigma_1 \sin(2\pi\phi),
\end{equation}
and
\begin{equation}
    K_p(\phi) = K_{p,0} + K_{p,1} \sin(2\pi\phi).
\end{equation}

The sinusoidal terms are included as a first-order test for phase-dependent structure across the transit. Because the transit occurs symmetrically around orbital phase $\phi = 0$, the function $\sin(2\pi\phi)$ acts as an odd (anti-symmetric) function over the narrow transit window, introducing a monotonic gradient across the transit chord. In this parameterisation, $A_0$ represents the mean absorption amplitude, while $A_1\sin(2\pi\phi)$ tests for an ingress-to-egress amplitude gradient, such as a leading--trailing asymmetry in the atmospheric absorption. Similarly, $\sigma_1\sin(2\pi\phi)$ and $K_{p,1}\sin(2\pi\phi)$ allow the CCF width and orbital velocity to vary with orbital phase, respectively. The normal priors centred on zero for these time-dependent terms allow the data to test whether such phase dependence is required.

We fit the model using a No-U-Turn Sampler (NUTS) implemented with \texttt{NumPyro} and \texttt{JAX}. Each fit is run for 2,000 iterations, with the first 1,000 as burn-in. Uniform priors are adopted for most parameters, while normal priors centred at zero are used for time-varying components ($A_1$, $K_{p,1}$, and $\sigma_1$). This choice enables a hypothesis test for whether any time-dependence is present.

\begin{table}[ht]
    \centering
    \caption{Model parameters and priors used in the forward modelling of the time-resolved cross-correlation function.}
    \label{tab:ccf_priors}
    \tiny
    \begin{tabular}{lll}
        \hline\hline
        Parameter & Description & Prior \\ \hline
        $A_0$ & Mean CCF amplitude & $\mathcal{U}(-1000,\,0)$ \\
        $A_1$ & Phase-dependent amplitude & $\mathcal{N}(0,\,0.1)$ \\
        $V_{\mathrm{sys}}$ & Systemic velocity & $\mathcal{U}(-35,\,-15)$ \\
        $K_p$ & Velocity gradient & $\mathcal{U}(200,\,350)$ \\
        $K_{p,1}$ & Phase-dependent velocity gradient & $\mathcal{N}(0,\,30)$ \\
        $\sigma_0$ & Mean CCF width & $\mathcal{U}(2,\,20)$ \\
        $\sigma_1$ & Phase-dependent width & $\mathcal{N}(0,\,30)$ \\
        \hline
    \end{tabular}
    \tablefoot{
        Uniform ($\mathcal{U}$) and normal ($\mathcal{N}$) priors are defined by their bounds and by their mean and standard deviation, respectively.
        Time-dependent parameters are assigned normal priors centred at zero to enable hypothesis testing for phase variability.
        All velocities and widths are expressed in km\,s$^{-1}$.
    }
\end{table}

To avoid underestimating uncertainties due to velocity-axis correlation in the CCF, we do not derive parameter errors directly from the MCMC posterior of a single fit. Instead, we adopt the bootstrapping approach we have developed: the individual flux values in the original spectra are jittered within their uncertainties, assuming Poisson statistics, and the entire downstream cross-correlation and forward-modelling procedure is repeated for each realisation. This process yields a distribution of best-fitting parameter values across the bootstrap trials, from which we estimate credible intervals.

As a validation of this choice, we compare the direct log-likelihood/MCMC posterior for the Fe\,II CCF from the first HARPS-N night with the corresponding bootstrap distribution (see Table~\ref{tab:feii_loglikelihood_bootstrap} in Appendix~D for the results of this comparison).

\subsection{Calculating weights for each observation}
\label{sect:weights}
To combine all transit observations in a statistically consistent manner, we assign a scalar weight to each night and species that reflects the expected contribution to the final detection. We derive these weights directly in cross-correlation function (CCF) space from four ingredients: (i) the significance of the amplitude from the two-dimensional fit to the CCF of the data; (ii) the significance of the amplitude from the two-dimensional fit to the injected model; (iii) the fraction of the transit that was actually observed; and (iv) the estimated noise level of the CCF continuum.

For each night, we fit the in-transit CCF stack with a simplified model as described in Sect.~\ref{sect:fitting_the_ccf}, with no time-dependent components for the amplitude, width, or $K_p$, and we extract the $z$-score associated with the fitted amplitude of the detection. Fitting the amplitude in this way ensures that the weight is evaluated only using the wavelength values contributing to the CCF. Additionally, we compare the detected signal to the expected signal when fitting the injected model and derive the corresponding statistical significance. We recognise that this approach depends on the specific model used; hence, we inject scaled MANTIS templates~\citep{Kitzmann_2021} as a standardised reference, thereby making any model dependence explicit and reproducible. This provides an estimate of the expected signal strength based on the spectrograph's wavelength coverage and naturally incorporates the data processing steps. Moreover, since some of our transit observations are incomplete, we weight each night by the fraction of the transit observed to account for the reduced information content. Lastly, we include the inverse of the measured noise level of the CCF continuum, which varies between nights and instruments depending on the spectrograph's overall noise characteristics.

\begin{equation}
    w \propto z_{\mathrm{CCF}} \times z_{\mathrm{inj}} \times (\Delta \phi_{\mathrm{transit}}) \times \frac{1}{\sigma_{\mathrm{cont}}},\label{eq:weight_eq}
\end{equation}

\noindent
where the weights are normalised over all nights to sum to unity.

\subsection{Detection significance}
\label{sect:detection_significance}
To quantify detection significance, we calculate a Mahalanobis $z$-score \citep{Mahalanobis_1936} that accounts for both amplitude parameters jointly. Since both $A_0$ and $A_1$ influence the absorption depth of the cross-correlation trace, we consider the null hypothesis where $A_0 = A_1 = 0$, corresponding to no signal. The detection significance is then defined as the Mahalanobis distance from the origin in the $(A_0, A_1)$ parameter space:

\begin{equation}
z = \sqrt{ \left( \begin{bmatrix} A_0 \\ A_1 \end{bmatrix} \right)^T \Sigma^{-1} \left( \begin{bmatrix} A_0 \\ A_1 \end{bmatrix} \right) }\label{eq:detection_stat}
\end{equation}

where $\Sigma$ is the covariance matrix of $A_0$ and $A_1$ derived from the MCMC posterior. This metric provides a robust measure of detection significance that incorporates uncertainty in both amplitude terms and their covariance, rather than assessing them independently.

We convert the z-score to a one-sided p-value under the assumption that absorption corresponds to negative amplitudes. Rather than impose a hard threshold for detection, we adopt a more transparent reporting strategy: we present the strength of evidence for each species as measured by its statistical significance, allowing for a graded interpretation of the results. This approach avoids the arbitrariness of fixed cutoffs, and supports reproducibility.

\section{Results}
We present our findings for the \vsys\, and $v_{\mathrm{rot}}\sin i$ of KELT-9A, and share results from applying the cross-correlation technique to detect gas species and resolve their Doppler motions in the atmosphere of KELT-9b from our bootstrapped, multi-instrument stacked CCF.

\begin{figure}[htbp]
  \centering
  \includegraphics[width=0.9\columnwidth, height=0.85\textheight, keepaspectratio]{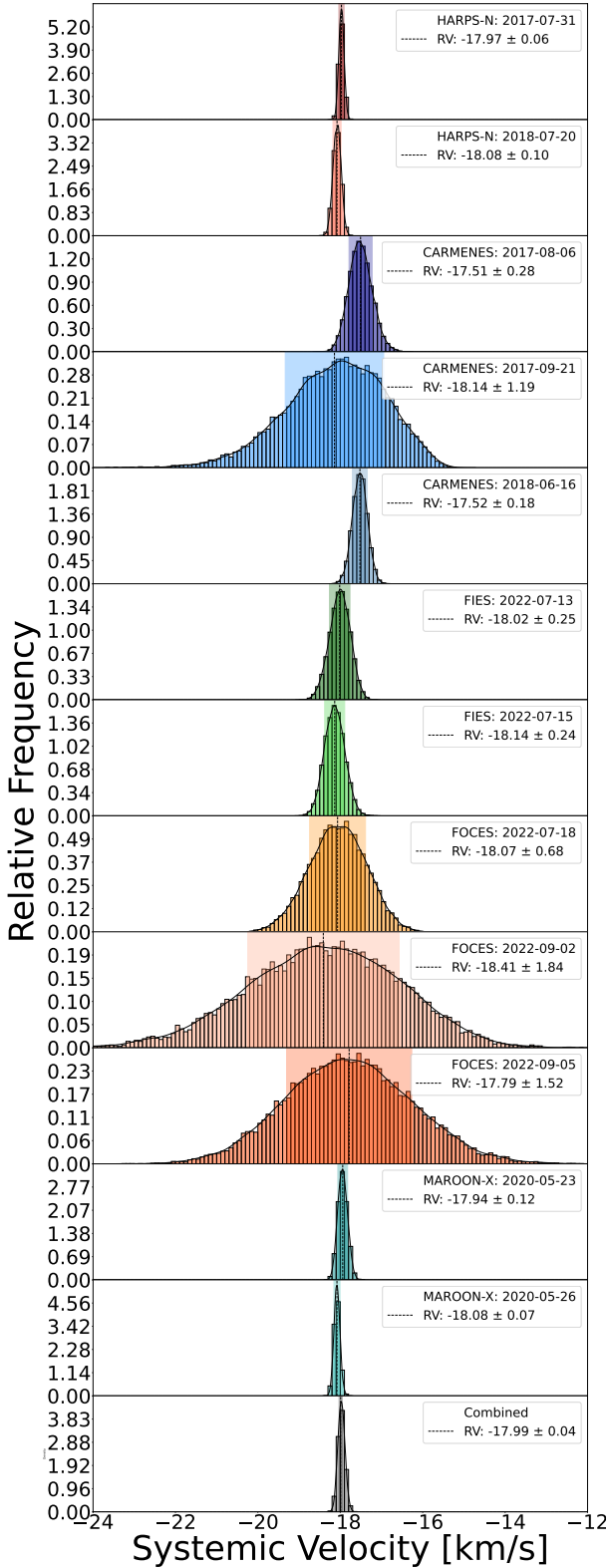}
  \caption{Distribution of measured systemic velocities for the out-of-transit exposures of KELT-9A. Each panel displays the radial velocity distribution of the CCF peak after correcting for the barycentric Earth radial velocity, normalised to unit area. The black dashed line indicates the mean systemic velocity for each night, and the shaded region shows the 1$\sigma$ uncertainty. The final panel (bottom) presents the combined distribution across all nights.}
  \label{fig:vsys_measurements}
\end{figure}

\begin{figure*}[htbp]
    \centering
    \includegraphics[width=0.95\textwidth,height=0.9\textheight,keepaspectratio]{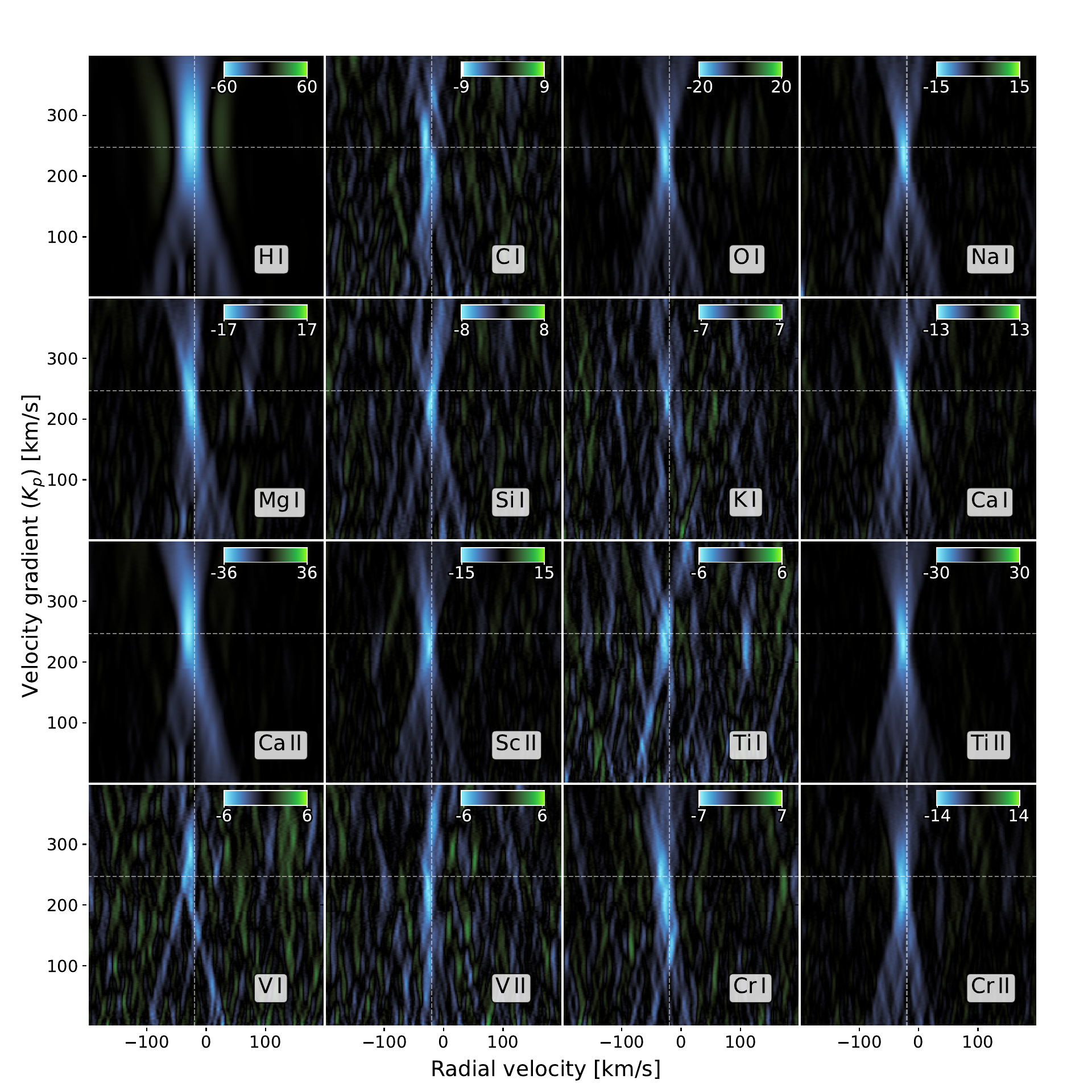}
    \caption{
    Velocity–velocity ($K_p$, $V_{\rm sys}$) maps of the gas species analysed in this study are shown here. Each panel presents the two-dimensional cross-correlation function (CCF), where colours denote signal strength in standard deviation units. Dashed lines mark the expected orbital velocity and systemic velocity of the system and planet. The species are arranged in order of increasing atomic number. For readability, the detections are split across two figure panels; this panel shows H I, C I, O I, Na I, Mg I, Si I, K I, Ca I, Ca II, Sc II, Ti I, Ti II, V I, V II, Cr I, and Cr II.
    }
    \label{fig:ccf_mosaic}
\end{figure*}

\begin{figure*}[htbp]
    \centering
    \includegraphics[width=0.95\textwidth,height=0.9\textheight,keepaspectratio]{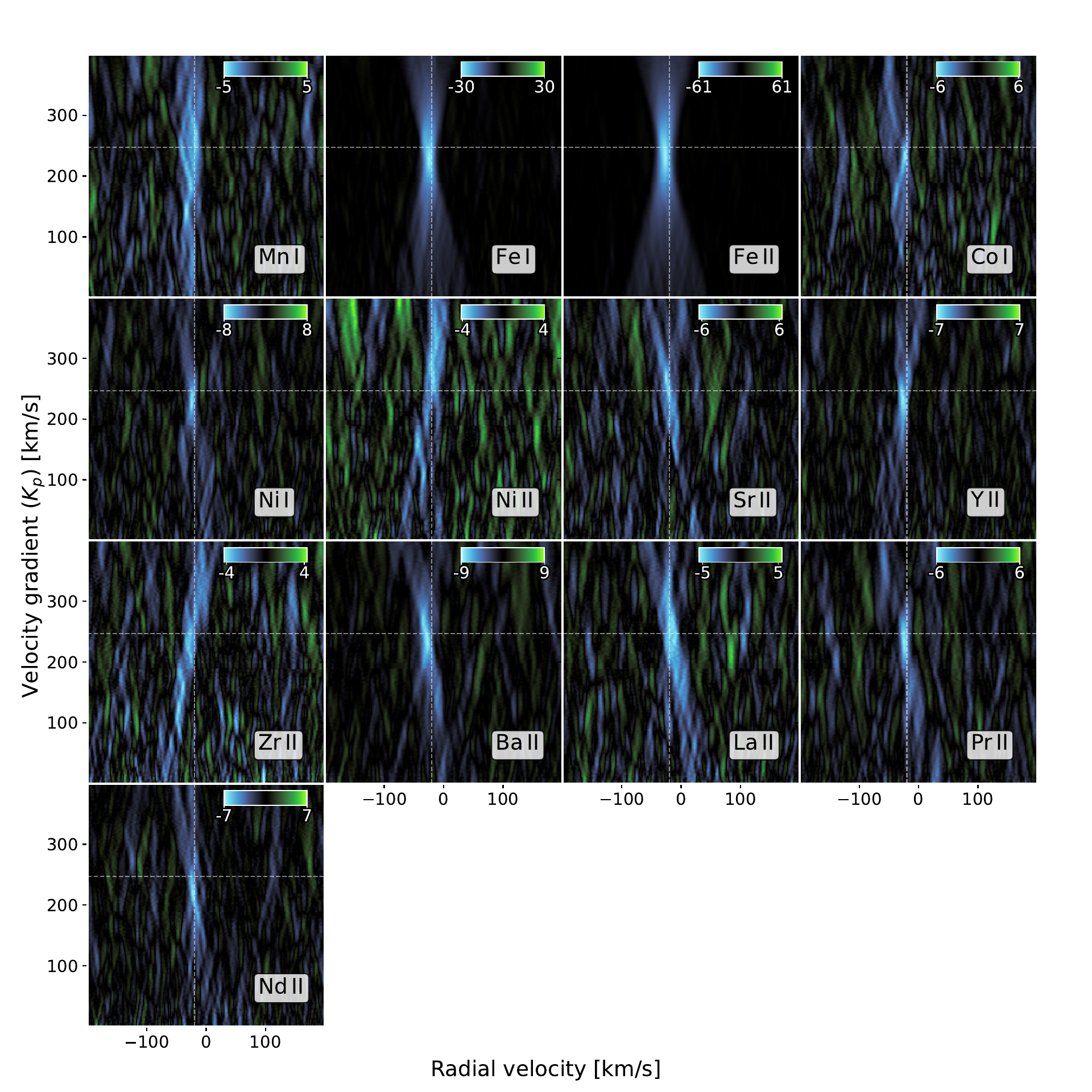}
    \caption{
    Continued velocity–velocity ($K_p$, $V_{\rm sys}$) maps for the remaining gas species analysed in this study. The plotting conventions are the same as in Fig.~\ref{fig:ccf_mosaic}. This panel shows Mn I, Fe I, Fe II, Co I, Ni I, Ni II, Sr II, Y II, Zr II, Ba II, La II, Pr II, and Nd II.
    }
    \label{fig:ccf_mosaic_part2}
\end{figure*}

\subsection{Systemic and rotational velocities of KELT-9A}
Our measurements of the \vsys\, of the KELT-9A system show consistent results across all nights, with overlapping distributions (Fig.~\ref{fig:vsys_measurements}). Although broad and shallow spectral lines can make precise radial velocity measurements challenging with standard Gaussian fits, our LSD approach appears to have overcome these issues, allowing for reliable, consistent measurements. There are no signs of systematic offsets between different nights or instruments; however, nights with lower S/N show broader distributions but still align with the overall value within uncertainties. Combining data from all 13 nights, the mean of the aggregate uncertainty-propagated distribution is $V_{\mathrm{sys}} = -17.99 \pm 0.04$\,km\,s$^{-1}$. Table~\ref{tab:vsys_stats_thiswork} provides the measured values per night, including the standard deviation and error. Our findings are in agreement with several published values, summarised in Table~\ref{tab:vsys_combined_literature}. The literature shows a range of systemic velocity measurements for KELT-9A, some based on HARPS-N data analysed with LSD~\citep{Borsa_2019,Asnodkar_2022b}, and yield differing results. This underscores the challenge of accurately measuring systemic velocities in rapidly rotating stars like KELT-9A.

\begin{table}[ht]
  \centering
  \scriptsize
  \setlength{\tabcolsep}{4pt}
  \renewcommand{\arraystretch}{0.9}
  \caption{Systemic velocity statistics for individual observing nights and for the combined dataset.}
  \label{tab:vsys_stats_thiswork}
  \begin{tabularx}{\columnwidth}{l X c c c}
    \hline\hline
    Instrument & Date & Mean & Std. dev. & Std. error \\
     &  & [km\,s$^{-1}$] & [km\,s$^{-1}$] & [km\,s$^{-1}$] \\
    \hline
    HARPS-N   & 2017-07-31 & $-17.973$ & $0.310$ & $0.060$ \\
    HARPS-N   & 2018-07-20 & $-18.080$ & $0.460$ & $0.097$ \\
    CARMENES  & 2017-08-06 & $-17.508$ & $1.310$ & $0.279$ \\
    CARMENES  & 2017-09-21 & $-17.484$ & $3.880$ & $1.076$ \\
    CARMENES  & 2018-06-16 & $-17.529$ & $1.340$ & $0.187$ \\
    FIES      & 2022-07-13 & $-18.210$ & $0.820$ & $0.239$ \\
    FIES      & 2022-07-15 & $-18.636$ & $0.910$ & $0.240$ \\
    FOCES     & 2022-07-18 & $-18.481$ & $1.530$ & $0.625$ \\
    FOCES     & 2022-09-02 & $-18.413$ & $4.470$ & $1.825$ \\
    FOCES     & 2022-09-05 & $-17.786$ & $5.470$ & $1.505$ \\
    MAROON-X  & 2020-05-23 & $-17.942$ & $0.390$ & $0.117$ \\
    MAROON-X  & 2020-05-26 & $-18.081$ & $0.320$ & $0.072$ \\
    \hline
    Combined & -- & $-17.992$ & $1.483$ & $0.039$ \\
    \hline
  \end{tabularx}
  \tablefoot{
    Mean, standard deviation, and standard error are computed from the posterior distributions of the systemic velocity for each observing night.
    The combined row corresponds to the aggregate distribution obtained by merging all nights.
  }
\end{table}

\begin{table}[ht]
  \centering
  \scriptsize
  \setlength{\tabcolsep}{4pt}
  \caption{Systemic velocity measurements from this work and from the literature.}
  \label{tab:vsys_combined_literature}
  \begin{tabularx}{\columnwidth}{X l l}
    \hline\hline
    Dataset / Source       & \vsys & Reference \\
     & [km\,s$^{-1}$] & \\
    \hline
    Combined (this work)   & $-17.992 \pm 0.04$ & This work \\
    PEPSI                  & $-17.860 \pm 0.044$ & \cite{Asnodkar_2022b} \\
    HARPS-N                & $-17.15 \pm 0.11$ & \cite{Asnodkar_2022b} \\
    HARPS-N                & $-17.74 \pm 0.11$ & \cite{Hoeijmakers_2019} \\
    HARPS-N                & $-19.819 \pm 0.024$ & \cite{Borsa_2019} \\
    TRES                   & $-18.97 \pm 0.12$ & \cite{Asnodkar_2022b} \\
    TRES                   & $-20.6 \pm 0.1$ & \cite{Gaudi_2017} \\
    GAIA                   & $-20.23 \pm 0.49$ & \cite{GAIA_1,GAIA_2} \\
    \hline
  \end{tabularx}
  \tablefoot{
    Uncertainties correspond to the quoted $1\sigma$ errors reported in each study.
    GAIA values are derived from astrometric solutions.
  }
\end{table}
\begin{table}[ht]
  \centering
  \small
  \setlength{\tabcolsep}{4pt}

  \caption{Projected rotational velocity statistics for individual spectrographs and for the combined dataset.}
  \label{tab:vsini_stats_thiswork}
  \begin{tabular}{l c c c}
    \hline\hline
    Instrument & Mean & Std. dev. & Std. error \\
     & [km\,s$^{-1}$] & [km\,s$^{-1}$] & [km\,s$^{-1}$] \\
    \hline
    HARPS-N   & $110.728$ & $0.285$ & $0.040$ \\
    CARMENES  & $108.333$ & $2.376$ & $0.144$ \\
    FIES      & $110.726$ & $0.755$ & $0.148$ \\
    FOCES     & $113.039$ & $3.160$ & $0.418$ \\
    MAROON-X  & $110.200$ & $0.300$ & $0.052$ \\
    \hline
    Combined & $110.538$ & $5.684$ & $0.054$ \\
    \hline
  \end{tabular}
  \tablefoot{
    Mean, standard deviation, and standard error are computed from the posterior distributions of $v\sin i$ for each instrument.
    The combined row corresponds to the aggregate multi-instrument distribution.
  }
\end{table}

The projected rotational velocity inferred from our LSD profile fits is stable across instruments and nights, as summarised in Table~\ref{tab:vsini_stats_thiswork}. Combining all nights, we obtain \vsini $= 110.54 \pm 0.05$\,km\,s$^{-1}$, and we place this in the broader literature context in Table~\ref{tab:vsini_combined_literature}. Our value agrees with the \vsini measurements reported by \citet{Gaudi_2017} and \citet{Borsa_2019} within their quoted uncertainties, and remains compatible with the somewhat higher estimate from \citet{Wyttenbach_2020} once its larger error bar is taken into account. It also overlaps with the GRACES-based measurement of \citet{Kama_2023}.

\begin{table}[ht]
  \centering
  \scriptsize
  \setlength{\tabcolsep}{4pt}
  \caption{Projected rotational velocity (\vsini) measurements from this work and from the literature.}
  \label{tab:vsini_combined_literature}
  \begin{tabularx}{\columnwidth}{X l l}
    \hline\hline
    Dataset / Source       & \vsini & Reference \\
                           & [km\,s$^{-1}$] & \\
    \hline
    Combined (this work)   & $110.538 \pm 0.054$ & This work \\
    TRES & $111.4 \pm 1.3$   & \cite{Gaudi_2017} \\
    HARPS-N & $111.8 \pm 1.0$   & \cite{Borsa_2019} \\
    HARPS-N & $116.9 \pm 1.8$   & \cite{Wyttenbach_2020} \\
    GRACES & $114.9 \pm 3.4$   & \cite{Kama_2023} \\
    \hline
  \end{tabularx}
  \tablefoot{
    Uncertainties correspond to the quoted $1\sigma$ errors reported in each study.
  }
\end{table}

\subsection{Detected atomic and ionised species}
\begin{figure*}
    \sidecaption
    \includegraphics[width=12cm]{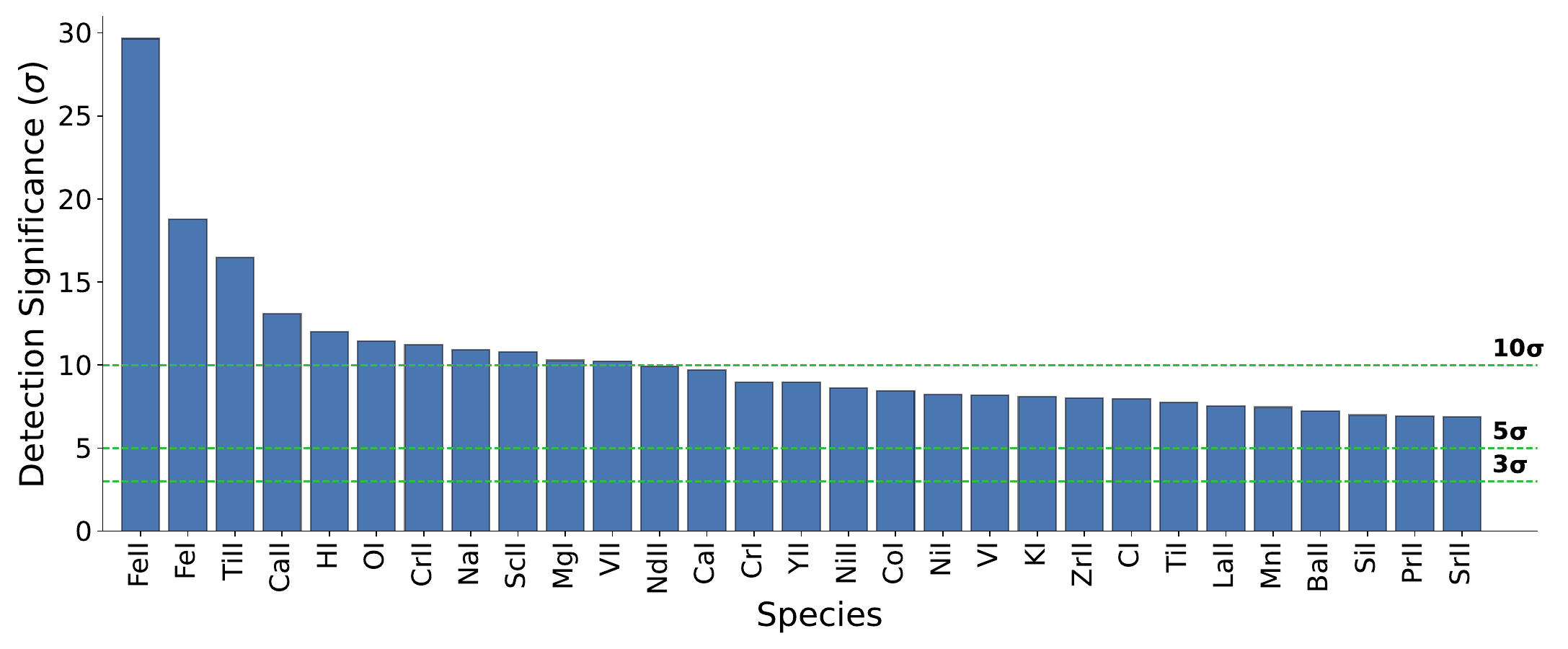}
    \caption{
    Detection significance ($z$-score) for each detected species, based on the Mahalanobis distance computed from the fitted amplitudes $A_0$ and $A_{1t}$. Horizontal dashed lines indicate conventional significance thresholds at $3\sigma$, $5\sigma$, and $10\sigma$. Species are ordered by descending detection strength as computed from the combined bootstrap posterior distributions.
    }
    \label{fig:zscore_species_barplot}
\end{figure*}

In this section, we present the cross-correlation results of an attempt to resolve the gas species in KELT-9b's atmosphere. Figs.~\ref{fig:ccf_mosaic} and \ref{fig:ccf_mosaic_part2} show the velocity-velocity maps for 29 atomic and ionised species in KELT-9b's atmosphere, which appear to display coherent signals very close to the systemic velocity and the orbital velocity of the planet, scaled to the number of standard deviations of the colour map. Our Mahalanobis $z$-scores are summarised in Fig.~\ref{fig:zscore_species_barplot} and detailed in Table~\ref{tab:zscores_species}, spanning over four orders of magnitude in one-sided $p$-value. We recover all previously reported species detections from high-resolution studies of KELT-9b \citep{Hoeijmakers_2018,Yan_2018,Yan_2019,Hoeijmakers_2019,Bello_Arufe_2022,Lowson_2023,Borsato_2023,Borsato_2024,DArpa_2024,Stangret_2024}, except Tb\,II.

We note, however, that not all velocity--velocity maps are characterised by a single compact peak. Several detections show broader, elongated, or locally structured absorption around the expected planetary velocities. We therefore avoid assigning a physical interpretation to these morphological details from the velocity--velocity maps alone. Instead, we treat these maps as qualitative diagnostics and rely on the bootstrap/Mahalanobis framework for the detection census.

The combined analysis detects 29 species, including additional absorbers C\,I, Si\,I, K\,I, V\,II, Co\,I, Zr\,II, La\,II, Pr\,II, and Nd\,II. All 29 recovered species exceed the conventional $5\sigma$ threshold in the bootstrap Mahalanobis significance test, with the weakest detection being Sr\,II at $z = 6.89$. Some of these species absorb strongly in the near-UV, and the evidence for their presence likely stems from the inclusion of the HIRES dataset, which extends down to 316\,nm. Additionally, we checked for aliasing by creating alias profiles, cross-correlating each template against the full set of other species considered in this study, to verify that the recovered signals are unlikely to be caused by blended absorption lines \citep{Borsato_2023}.

\begin{figure*}[htbp]
    \centering
    \includegraphics[width=\textwidth]{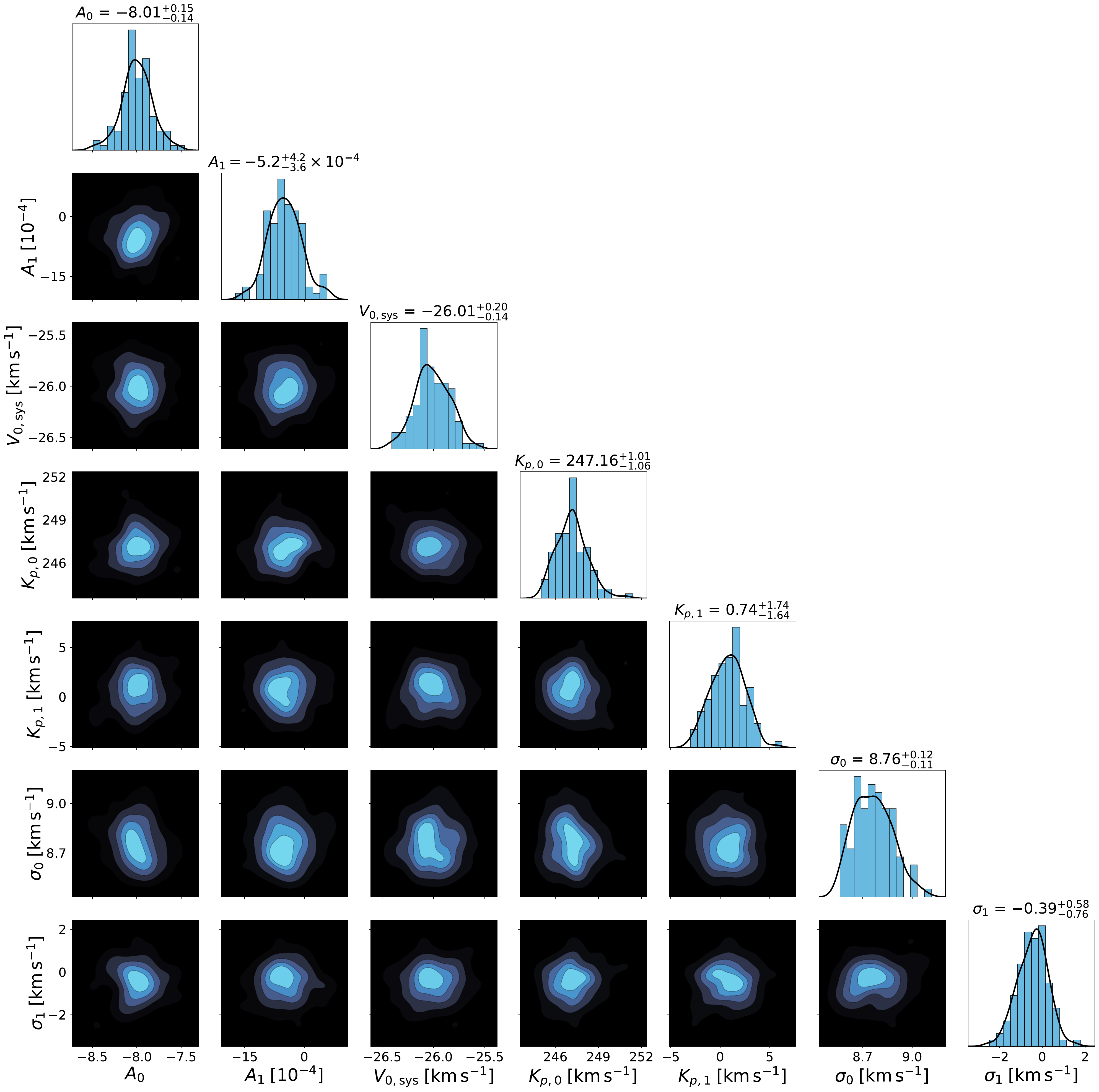}
    \caption{
    Corner plot showing the distribution and pairwise covariances of best-fit parameters retrieved from the full census of statistically significant species detections. The parameters include the constant and time-dependent amplitudes ($A_0$, $A_1$), systemic velocity (\vsys), orbital semi-amplitude and its phase variation ($K_p$, $K_{p,1}$), and Gaussian widths ($\sigma_0$, $\sigma_1$). One-dimensional histograms along the diagonal show the distribution of each parameter, while the off-diagonal panels reveal correlations across the ensemble of species. This figure summarises the diversity of line profiles and kinematic structures observed across the atmosphere of KELT-9b.
    }
    \label{fig:corner_plot}
\end{figure*}

\subsection{A census of the retrieved parameters}
We aggregated the posterior distributions of each detection to obtain a census of the retrieved cross-correlation amplitude, radial velocity offset, velocity gradient, and line width for all detections presented in this study. Fig.~\ref{fig:corner_plot} visualises the distribution and covariances of these global model parameters, derived from the census of all significant detections. The constant and time-dependent components of each parameter, along with their median values and credible intervals, are summarised in Table~\ref{tab:posterior_summary}. This population-level posterior view is the relevant summary for the census analysis presented here. The full set of individual species posteriors, together with their detailed covariance structure and physical interpretation, is deferred to future work, where those species-by-species comparisons can be treated explicitly without shifting the focus of the present paper away from the methodological framework and the global atmospheric census. The population-level results nevertheless reveal several notable features. The uncertainties quoted for these global census parameters should therefore be interpreted as population-level uncertainties, not as the formal precision of any single high-S/N species. They combine the bootstrap variance of the fitted CCF parameters with the dispersion among detected species, and therefore include real or apparent species-to-species heterogeneity in the recovered atmospheric velocities. This conservative treatment avoids forcing all tracers to share a single sharply defined $K_p$ or $V_{0,\mathrm{sys}}$, since different species may probe different pressures, altitudes, line strengths, or dynamical regions of the atmosphere.

The absorption amplitude shows no significant dependence on time, with the variable component centred near zero ($A_{1} = (-5.2 \pm 4.1) \times 10^{-4}$). The radial velocity value, $V_{0,\mathrm{sys}} = -26.01 \pm 0.17$\,km\,s$^{-1}$, is notably blue-shifted compared to the measured stellar systemic velocity of $V_{\mathrm{sys}} = -17.99 \pm 0.04$\,km\,s$^{-1}$, corresponding to an offset of $\Delta V_{\mathrm{offset}} = -8.02 \pm 0.17$\,km\,s$^{-1}$. The distribution of retrieved orbital velocities peaks around $K_{p,0} = 247.16 \pm 1.06$\,km\,s$^{-1}$, consistent with the expected Keplerian value of $247.4 \pm 5.7\,\mathrm{km\,s^{-1}}$.\footnote{Derived using $a = 0.03368 \pm 0.00078\,AU$ \citep{Borsa_2019} and $P = 1.48111874 \pm 0.00000014\,d$ \citep{Kokori_2023}} The time-dependent velocity and width components remain consistent with zero ($K_{p,1} = 0.74 \pm 1.65$\,km\,s$^{-1}$; $\sigma_{1} = -0.39 \pm 0.69$\,km\,s$^{-1}$), with the constant line width well constrained at $\sigma_0 = 8.76 \pm 0.11$\,km\,s$^{-1}$.

\begin{table}[ht]
\centering
\caption{Bootstrap estimates for the global model parameters.}
\begin{tabular}{lr}
\hline
Parameter & Bootstrap value \\
\hline
$A_0$ & $-8.005 \pm 0.171$ \\
$A_{1}$ & $-0.00052 \pm 0.00041$ \\
$V_{0,\mathrm{sys}}$ [km\,s$^{-1}$] & $-26.014 \pm 0.166$ \\
$K_{p,0}$ [km\,s$^{-1}$] & $247.156 \pm 1.058$ \\
$K_{p,1}$ [km\,s$^{-1}$] & $0.742 \pm 1.646$ \\
$\sigma_0$ [km\,s$^{-1}$] & $8.759 \pm 0.113$ \\
$\sigma_{1}$ [km\,s$^{-1}$] & $-0.385 \pm 0.689$ \\
\hline
\end{tabular}
\tablefoot{Bootstrap values correspond to the median (50th percentile) of the 100 combined bootstrap realisations, with uncertainties given by their standard deviations.
These global uncertainties describe the repeatability of the weighted census result under the flux-level bootstrap, rather than the formal uncertainty of an individual species fit.
Amplitudes $A_0$ and $A_1$ are dimensionless and are expressed in units of the standard deviation of the cross-correlation function.}
\label{tab:posterior_summary}
\end{table}

\section{Discussion}
This section examines the implications of the radial, and rotational velocity measurements of the KELT-9 system. We also interpret atmospheric detections over all 13 nights of high-resolution transit observations of KELT-9b using a census-based approach. This is enabled by our bootstrapping method, which accounts for the velocity-axis correlation caused by cross-correlation with fine velocity steps. Our comprehensive dataset allows us to move beyond individual detections and analyse the population-level behaviour of key physical parameters. We assess whether the gas species' combined population shows time variation in the transit signature. Overall, this initial finding provides a clear overview of the average motion of the strongest absorbing species in KELT-9b’s atmosphere, paving the way for a more detailed, species-specific analysis in future work.

\subsection{Systemic and rotational velocities of KELT-9A}
We briefly outline our efforts to determine the systemic velocity of the KELT-9A system and the star's vsini. Measuring the systemic velocity for a fast-rotating star like KELT-9A is challenging because of the broad stellar lines caused by rapid rotation. At such broadening levels, the stellar lines are susceptible to distortion over wide wavelengths they span, reducing the precision of standard radial-velocity extraction techniques \citep{Bouchy_2001,Borsa_2019}. Factors like instrumental artefacts, atmospheric changes, and imperfect continuum normalisation can also distort the cross-correlation profile and lead to biased centroid estimates~\citep{Malavolta_2017,Bechter_2021,Ivanova_2023}. Despite these issues, our results show good agreement across observing nights (see Table~\ref{tab:vsys_stats_thiswork}), with dispersions of 1--2\,km\,s$^{-1}$.

Throughout this study, we report both the standard deviation and the standard error of the mean (SEM) of the systemic-velocity measurements, as they serve complementary roles (Table~\ref{tab:vsys_stats_thiswork}). The standard deviation captures the total variance in the posterior distribution—accounting for measurement noise and modelling limitations—while the SEM reflects the internal precision in estimating the mean. Thus, the SEM should not be viewed as the uncertainty of an absolute systemic velocity, but is included here to demonstrate the potential precision of our method with this dataset. When all nights are combined, the aggregated bootstrap distribution yields a highly precise mean with an uncertainty of 0.04\,km\,s$^{-1}$ (Table~\ref{tab:vsys_combined_literature}). This 0.04\,km\,s$^{-1}$ uncertainty represents the precision of the aggregate mean, while the 1.48\,km\,s$^{-1}$ standard deviation of the combined distribution captures the broader night-to-night and instrument-to-instrument reproducibility.

Previous studies have employed least-squares deconvolution (LSD) to extract average line profiles from high-resolution spectra, reducing noise and enhancing stability \citep{Borsa_2019,Asnodkar_2022b}. The published radial-velocity values for KELT-9A show some variation between analyses of similar or overlapping datasets (Table~\ref{tab:vsys_combined_literature}).

We comment on the projected stellar rotational velocity, \vsini, which was estimated from the widths of the LSD profiles. Unlike the systemic velocity, which depends on measuring small shifts in the centroid of very broad lines, \vsini\, is primarily determined by the overall profile broadening and remains consistent across instruments and nights (Table~\ref{tab:vsini_stats_thiswork}). Combining all nights yields \vsini$ = 110.54 \pm 0.05$\,km\,s$^{-1}$. The literature context is summarised in Table~\ref{tab:vsini_combined_literature}. This value is consistent with previous findings by \cite{Gaudi_2017}, \cite{Borsa_2019}, and \cite{Kama_2023}. These results indicate that the adopted rotational broadening scale is robust, even when the absolute systemic velocity is more challenging to constrain for such a rapidly rotating star.

\subsection{Bootstrapping for cross-correlation analyses}
In high-resolution transmission spectroscopy, bootstrapping provides a natural way to propagate data uncertainties through the full sequence of processing steps in cross-correlation analyses. This is particularly relevant in CCF studies, where several effects introduce correlations along the velocity axis, including instrumental resolution, Doppler smearing from the planet's varying radial velocity, and interpolation required to align the flux on a common velocity grid. Additionally, processing in the wavelength domain, including continuum normalisation and telluric correction with Molecfit, can distribute information across pixels. By resampling the data and rerunning the analysis pipeline, bootstrapping eliminates the need to explicitly model these effects at the CCF likelihood level, thereby capturing the influence of correlated structures on the resulting uncertainty estimates.

Alternative approaches are, however, possible. Methods that fit the flux data directly, such as those commonly employed with high-resolution atmospheric retrievals, avoid velocity-axis correlations introduced by velocity re-gridding~\citep[e.g.][]{Gibson_2020}. Similarly, if performed with care, it is likely possible to select a radial-velocity grid that yields reasonable uncertainty estimates; as shown in Fig.~\ref{fig:amp_unc_vs_step}, choosing a grid spacing close to the appropriate scale already produces a sensible uncertainty estimate, even if not perfectly matched, and may substantially simplify the fitting procedure, particularly in single-instrument analyses. In our case, however, where we aim to combine heterogeneous datasets spanning different exposure times and resolutions (and therefore different levels of Doppler smearing; Sect.~\ref{sect:datasets}), the ability of bootstrapping to propagate uncertainties consistently while adopting a single common velocity grid across all nights makes it the most appropriate and dependable approach for our science goals.

\subsection{The chemical inventory of KELT-9b and the census of the fitted parameters}
The combined analysis yields 29 statistically significant atmospheric detections, with the corresponding velocity--velocity maps shown in Figs.~\ref{fig:ccf_mosaic} and \ref{fig:ccf_mosaic_part2}. Several of these detections are facilitated by the near-UV coverage of the HIRES dataset. These results collectively indicate a diverse array of gaseous species that can be measured at high spectral resolution. In addition to compiling the species inventory, the modelling yields fitted parameter estimates for each detection. Although detailed interpretation is deferred to future work, the census presented here provides a consistent set of measurements across the detected species for population-level comparison.

The distribution of Mahalanobis $z$-scores provides a conservative statistical assessment of the recovered atmospheric inventory. Unlike a significance estimate based on a single local peak in a $K_p$--$V_{\mathrm{sys}}$ map, the adopted metric evaluates the joint fitted signal and its covariance within the bootstrap forward-model framework. The coherent structures seen in the velocity--velocity maps are therefore supported by a significance measure that incorporates the full transit model and propagated flux uncertainties. The recovery of both neutral and singly ionised refractory species is consistent with these absorbers being present in the observable atmosphere of KELT-9b, providing a basis for the more detailed chemical and dynamical analysis deferred to future work.

Radial velocity offsets from cross-correlation detections are sometimes discussed in terms of an inferred systemic velocity. What is recovered is a Doppler offset of the planetary absorption signal relative to the stellar rest frame. This quantity only reduces to \vsys\, in the special case where the atmosphere contributes no net line-of-sight motion, where absorption is not preferentially shifted to the blue or the red. For ultra-hot Jupiters, this condition is generally unmet. Strong day-to-night flows can produce a net blueshift in the transmission spectrum of several km s$^{-1}$ \citep[e.g.][]{Snellen_2010_CC,Louden_2015,Wyttenbach_2015,Brogi_2016,Seidel_2019,Ehrenreich_2020}. Furthermore, the magnitude of the shift may differ between species if distinct lines probe different atmospheric regions~\citep[e.g.][]{Merritt_2021,Kesseli_2022,Seidel_2025,Prinoth_2025}.

From the stellar lines we obtain $V_{\mathrm{sys}} = -17.99 \pm 0.04$\,km\,s$^{-1}$, while the global census fit returns a planetary offset of $V_{0,\mathrm{sys}}=-26.01 \pm 0.17$\,km\,s$^{-1}$ (Table~\ref{tab:posterior_summary}). The difference between these values corresponds to a net blueshift of $\Delta V_{\mathrm{offset}} = -8.02 \pm 0.17$\,km\,s$^{-1}$ in the planet frame, suggestive of a bulk day-to-night component when considering the median solution. The corresponding species-specific $V_{0,\mathrm{sys}}$ values are listed in Table~\ref{tab:v0_species}; these show that most species favour blueshifted offsets, although with scatter between species. Given the broad planetary posterior, this result is interpreted as an indicative global offset rather than a species-independent wind measurement. Furthermore, species-specific offsets provide a means to investigate the dynamical structure of the upper atmosphere when interpreted within a consistent stellar reference frame \citep[e.g.][]{Merritt_2021,Kesseli_2022,Seidel_2025,Prinoth_2025}.

One outcome of the global census fit is that the constant orbital-velocity parameter aligns closely with the expected Keplerian motion of the planet. We recover $K_{p,0}=247.16 \pm 1.06$\,km\,s$^{-1}$ (Table~\ref{tab:posterior_summary}), consistent with the calculated value of $247.4\pm5.7$\,km\,s$^{-1}$. Although individual species exhibit dispersion in their retrieved $K_p$ values from localized dynamical effects or differing line-formation depths \citep{Wardenier_2021,Prinoth_2022,Kesseli_2022}, these variations appear to cancel out across the 29-species ensemble, yielding no net orbital-velocity shift at the population level.

The global line width is similarly well constrained, with a median of $\sigma_0=8.76 \pm 0.11$\,km\,s$^{-1}$ (Table~\ref{tab:posterior_summary}). The spread in $\sigma_0$ is plausibly set via a blend of physical and instrumental broadening, including unresolved wind structure and the superposition of atmospheric regions with distinct line-of-sight velocities, convolved with the instrumental line-spread function \citep{Kempton_2014,Brogi_2016,Bourrier_2013,Oklo_Hirata_2018}.

Including time-dependent terms in the global model is a way to ask whether the planetary signal changes systematically across the transit. If the atmosphere is not longitudinally uniform, then one could in principle see trends in the absorption amplitude, the velocity trace, or the line width from ingress to egress, for example from leading--trailing limb asymmetries, a displaced hotspot, or wind patterns that shift the dominant absorbing region during the transit chord \citep{Knutson_2007,Showman_2011,Mikal_Evans_2023,Wardenier_2025}. At the population level, however, no statistically significant time dependence is recovered in any of these quantities. The amplitude term is centred on zero ($A_{1}=(-5.2 \pm 4.1) \times 10^{-4}$), and both the velocity-variation and width-evolution terms are likewise consistent with zero ($K_{p,1}=0.74 \pm 1.65$\,km\,s$^{-1}$; $\sigma_{1}=-0.39 \pm 0.69$\,km\,s$^{-1}$; Table~\ref{tab:posterior_summary}). These results imply that the stacked, species-agnostic signal is broadly stable across the transit, and that any phase dependence is either weak, averages out across the ensemble, or sits below the sensitivity of this census. This does not mean individual species cannot show time dependence: different neutral and ionised tracers can probe different regions and conditions in the atmosphere, and may therefore show species-specific behaviour that is washed out in the global fit \citep{Pino_2018,Stangret_2020,Kesseli_2022}. Assessing this properly requires a targeted species-by-species analysis.

\section{Conclusion}
Using 13 transit observations from six high-resolution spectrographs, we performed a global cross-correlation analysis of the ultra-hot Jupiter KELT-9b. This comprehensive approach yields 29 statistically significant detections in KELT-9b's atmosphere. These detections were established using a bootstrapped Mahalanobis-significance framework, providing a conservative measure of detection confidence.

We forward-modelled the full cross-correlation trace using a time-dependent Gaussian model, recovering a planetary velocity gradient consistent with the known orbital velocity, a systemic offset indicative of a net atmospheric blueshift, and stable line widths across transit. The time-dependent components remained consistent with zero.

Together, these findings offer a statistically anchored view of the bulk atmospheric structure and motion, while motivating more granular retrievals. Future work will build on this foundation to analyse species individually, where spatial, thermal, and chemical diversity may emerge with greater clarity.

\begin{acknowledgements}
This work has made use of the VALD database, operated at Uppsala University, the Institute of Astronomy RAS in Moscow, and the University of Vienna. This work has made use of data from the European Space Agency (ESA) mission Gaia (\url{https://www.cosmos.esa.int/gaia}), processed by the Gaia Data Processing and Analysis Consortium (DPAC, \url{https://www.cosmos.esa.int/web/gaia/dpac/consortium}). Funding for the DPAC  has been provided by national institutions, in particular the institutions  participating in the Gaia Multilateral Agreement. N.W.B.\ acknowledges funding from Agence Nationale pour la Recherche (ANR, project ANR-24-CE49-3397 ORVET). B.T.\ acknowledges the financial support from the Wenner-Gren Foundation (WGF2022-0041). S.P.\ acknowledges support from the Swiss National Science Foundation under grant 51NF40\_205606 within the framework of the National Centre of Competence in Research PlanetS.

\\\indent Part of the data presented herein were obtained at the W.\ M.\ Keck Observatory, which is operated as a scientific partnership among the California Institute of Technology, the University of California and the National Aeronautics and Space Administration. The Observatory was made possible by the generous financial support of the W.\ M.\ Keck Foundation. The authors wish to recognise and acknowledge the very significant cultural role and reverence that the summit of Maunakea has always had within the indigenous Hawaiian community. We are most fortunate to have the opportunity to conduct observations from this mountain.

\\\indent Based on observations made with the Italian Telescopio Nazionale Galileo (TNG), operated on the island of La Palma by the INAF – Fundación Galileo Galilei at the Roque de Los Muchachos Observatory of the Instituto de Astrofísica de Canarias (IAC); data from the CARMENES data archive at CAB (INTA-CSIC); and observations obtained with the MAROON-X instrument, developed by the University of Chicago, and installed at the Gemini North telescope. Part of the observations presented herein were obtained with the 2 m Fraunhofer telescope located at the Wendelstein Observatory (operated by the University Observatory Munich, Ludwig‑Maximilians‑Universität München) in the Bavarian Alps, Germany.

\\\indent Based on observations made with the Nordic Optical Telescope, owned in collaboration by the University of Turku and Aarhus University, and operated jointly by Aarhus University, the University of Turku and the University of Oslo, representing Denmark, Finland and Norway, the University of Iceland and Stockholm University at the Observatorio del Roque de los Muchachos, La Palma, Spain, of the Instituto de Astrofisica de Canarias. The NOT data were obtained under programme ID 65-001.

\end{acknowledgements}

\bibliographystyle{aa}
\bibliography{bib}

\begin{appendices}
\makeatletter
\@fleqnfalse
\makeatother

\section{Extended observational details}\label{sect:datasets}
This appendix presents extended technical details regarding the observations and data reduction procedures that support the main results. The weights applied to combine observations and a summary of bad pixel statistics are described.

\subsection{Weights for the cross-correlation function}\label{sect:appendix_weights}
When constructing the combined cross-correlation functions, we assign relative weights to each observing night on a species-by-species basis. These weights reflect the extent to which each night contributes to the final stacked signal for a given species, and naturally account for differences in wavelength coverage, signal-to-noise, and data quality between instruments and epochs. Table~\ref{tab:species_weights} lists the resulting normalised weights, where values closer to unity indicate nights that dominate the combined detection for that species, while smaller values indicate only a minor contribution.

\subsection{Summary of masked-pixel statistics}\label{app:masked_pixels}
This section presents the fraction of pixels masked during the preprocessing stage for each observing night. The values were obtained directly from the \texttt{tayph} pipeline and include both automatically flagged narrow artefacts and broader manually masked wavelength regions. The automatic isolated-pixel component is small for most datasets; the higher values in Table~\ref{tab:badpix_stats} are instead driven by broad wavelength/order exclusions associated with telluric residuals, low-throughput regions, or order-level extraction artefacts. Representative examples for the two highest-fraction cases, CARMENES and HIRES, are shown in Figs.~\ref{fig:carmenes_mask_diagnostic} and \ref{fig:hires_mask_diagnostic}. The CARMENES masked regions are dominated by unstable low-throughput behaviour at the blue edge and by deep telluric absorption in the redder orders. In the deepest telluric bands, producing a fit becomes difficult, so the saturated line cores or broader affected intervals were masked while shallower telluric regions were retained where the correction was adequate. The affected HIRES regions are dominated by order-level artefacts, including exposure-dependent quasi-periodic structure, glitch-like pixels, continuum/extraction artefacts, and flattened or clipped flux behaviour; representative examples are shown in Fig.~\ref{fig:hires_mask_diagnostic}. These regions produced unstable cross-correlation residuals when less aggressive masks were tested, so they were excluded conservatively to prevent artificial CCF structure from entering the atmospheric species census. Table~\ref{tab:badpix_stats} provides the total number of pixels per exposure and the proportion affected for each spectrograph and night.

\vspace{-\baselineskip}
\section{Telluric corrections regions}\label{sect:telluric_regions}
Table~\ref{tab:telluric_regions} lists the wavelength intervals (in µm) used for telluric fitting with Molecfit for each spectrograph. These regions cover \ce{H2O} and \ce{O2} absorption bands in areas where the continuum is well behaved. Minor adjustments were made when necessary on a per-night basis.

\vspace{-\baselineskip}
\section{Additional systemic velocity analysis}
This appendix provides supplementary context for the systemic-velocity measurements. Fig.~\ref{fig:EX_first_exposures_grid} displays the LSD profiles for the first out-of-transit exposure of each observing night, along with the best-fit rotational profile and corresponding residuals. The residuals do not exhibit significant systematic structure. Table~\ref{tab:vsys_by_spectrograph} summarises the systemic-velocity statistics by spectrograph, reporting the mean, standard deviation, and standard error of the mean. These results demonstrate that the inferred \vsys values are generally consistent across instruments.

\vspace{-\baselineskip}
\section{Additional cross-correlation results}
This section provides supplementary material detailing the cross-correlation cleaning steps and the statistical significance of the planetary detections. Fig.~\ref{fig:ccf_cleaning} illustrates the sequential application of Doppler shadow subtraction and vertical detrending on a representative HARPS-N dataset, as discussed in Sect.~\ref{sect:fitting_the_ccf}. Fig.~\ref{fig:ccf_cleaning} uses the Fe II CCF; the plotted values are scaled by subtracting the median CCF value measured in the velocity regions $[-200,-150]$ and $[150,200]$~km\,s$^{-1}$ and dividing by the standard deviation measured over the same regions. These preprocessing steps are crucial for isolating the planetary signal from stellar and instrumental noise.

Fig.~\ref{fig:negative_kp_injection_control} shows the corresponding negative-$K_p$ model-injection control for Fe\,I. The injected model was processed through the same Doppler-shadow removal, vertical detrending, cleaning, and weighted-combination steps as the science CCFs. The recovery of the injected feature on the negative-$K_p$ side demonstrates that the cleaning procedure does not erase a planetary-like signal with a different velocity trajectory or force it toward the expected positive-$K_p$ solution; the positive-$K_p$ Fe\,I signal remains visible in the combined map.

Table~\ref{tab:zscores_species} presents the Mahalanobis $z$-scores for each detected species, quantifying the detection significance derived from the joint amplitude distribution and serving as a census of all confidently retrieved atmospheric signals. We present a comparison between a direct log-likelihood/MCMC fit and our adopted bootstrap framework for Fe\,II in Table~\ref{tab:feii_loglikelihood_bootstrap}. For this validation, the direct MCMC fit uses the cleaned, non-bootstrap Fe\,II CCF from the first HARPS-N night, while the bootstrap values are derived from the distribution of fitted parameters across bootstrap realisations for the same night. The listed $z$-score is the Mahalanobis detection significance computed from the joint amplitude parameters, following Sect.~\ref{sect:detection_significance}. The species-census Mahalanobis $z$-scores reported elsewhere in Appendix~D refer to the combined multi-night dataset, and therefore should not be compared directly to the single-night Fe\,II validation $z$-scores listed in Table~\ref{tab:feii_loglikelihood_bootstrap}. The comparison shows that while the central kinematic parameters remain consistent between the two methods, the bootstrap framework yields substantially more conservative uncertainties. The corresponding species-specific systemic velocity offsets ($V_{0,\mathrm{sys}}$) derived from the bootstrap fits are summarised in Table~\ref{tab:v0_species}. Finally, Table~\ref{tab:ism_features} provides a summary of the likely interstellar medium absorption features identified across our observations.

For this single-night Fe\,II test case, the direct MCMC and bootstrap analyses recover consistent central values for the kinematic parameters and CCF widths, indicating that the adopted bootstrap procedure does not shift the inferred planetary signal. The main difference is in the uncertainty scale: the direct likelihood posterior is much narrower and yields a substantially larger Mahalanobis detection significance. This behaviour is consistent with the single-map likelihood not fully capturing correlations introduced by cross-correlation, interpolation, Doppler smearing, and the common velocity grid. We therefore retain the bootstrap-derived parameter uncertainties and detection significances as the conservative values used for the full species census.

\subsection{Polynomial-order diagnostics}
\label{app:poly_order_diagnostics}
To select the polynomial degree for vertical detrending, we perform an order-sweep test on both science CCFs and CCFs containing an injected forward-model signal. The planetary trail ($\pm 20\mathrm{\,km\,s^{-1}}$) is masked before fitting each velocity column. The left panel of Fig.~\ref{fig:r4_07_order_selection} shows the background RMS, computed as the standard deviation of cleaned CCF pixels in an off-trail velocity window ($100 \le |v| \le 300\mathrm{\,km\,s^{-1}}$). The middle panel shows the RMS distortion of the injected signal, calculated as the root-mean-square difference between the true injected model and the recovered model signal within the planet velocity window ($|v| \le 50\mathrm{\,km\,s^{-1}}$). The right panel shows their joint trade-off. The background RMS decreases rapidly up to degree 5 before flattening, whereas higher degrees progressively distort the injected planet signal. Degree 5 represents the optimal trade-off between noise removal and signal preservation. Table~\ref{tab:r4_07_order_metrics} lists the metrics for degrees 1--10.

\FloatBarrier

\vspace{-\baselineskip}
\section{Additional derivations for the velocity-axis correlation}
\label{sect:additional_velocity-axis_correlation}
\subsection{Estimating the interpolation noise when oversampling the radial velocity grid}
\label{app:interpolation_Noise}
In this section, we derive the estimated increase in uncertainty due to data interpolation. Formally, for correlated samples, the effective number of independent measurements $N_{\mathrm{eff}}$ can be approximated by

\begin{equation}
    N_{\mathrm{eff}} = \frac{N}{1 + 2\sum_{k=1}^{\infty}\rho(k)},
\end{equation}

\noindent
where $\rho(k)$ is the autocorrelation at lag $k$. For a discrete sequence, the autocorrelation at lag $k$ measures the correlation between values separated by $k$ sampling steps; a lag of $k=1$ therefore corresponds to correlations between adjacent velocity bins, while larger lags probe correlations over wider separations. To first order, this can be expressed in terms of the lag-1 correlation coefficient as

\begin{equation}
    N_{\mathrm{eff}} \approx N \frac{1 - \rho_1}{1 + \rho_1}.
\end{equation}

Since the uncertainty on a fitted parameter scales inversely with the square root of the effective sample size ($\sigma_A \propto N_{\mathrm{eff}}^{-1/2}$), the corresponding inflation in uncertainty is given by

\begin{equation}
    \frac{\sigma_{A,\mathrm{corr}}}{\sigma_{A,\mathrm{ideal}}} = \sqrt{\frac{N}{N_{\mathrm{eff}}}}
    = \sqrt{\frac{1 + \rho_1}{1 - \rho_1}}.
\end{equation}

For $\rho_1 \simeq 0.1$, this predicts an inflation factor of $\sim1.09$, which appears to account for the offset between the bootstrap and analytic estimates in Fig.~\ref{fig:amp_unc_vs_step}.

\subsection{Amplitude–width degeneracy and its contribution to the uncertainty inflation}
\label{app:amplitude_degeneracy}
In addition to the information loss caused by coarse sampling, an additional source of uncertainty is caused by the covariance between the amplitude $A$ and width $\sigma$ parameters of the fitted Gaussian profile. When the sampling interval $\Delta v$ approaches a non-negligible fraction of the intrinsic width $\sigma$, the degeneracy between $A$ and $\sigma$ increases: a slightly broader profile can reproduce the same integrated line area with a smaller amplitude. This parameter coupling inflates the marginal uncertainty of $A$ beyond the Fisher-information limit derived for fixed $\sigma$.

To quantify this effect, we consider the Fisher Information Matrix (FIM) for a Gaussian model with independent Gaussian noise of variance $\sigma_{\mathrm{noise}}^2$. For a forward model $f_i(\boldsymbol{\theta})$ evaluated at velocities $v_i$, the Fisher Information Matrix is defined as~\citep{Cramer_1946,Rao_1945}

\begin{equation}
    \mathcal{I}_{mn}
    =
    \frac{1}{\sigma_{\mathrm{noise}}^2}
    \sum_i
    \frac{\partial f_i}{\partial \theta_m}
    \frac{\partial f_i}{\partial \theta_n},
\end{equation}

\noindent
where $\boldsymbol{\theta}$ denotes the set of model parameters and the indices $m,n$ run over those parameters. The diagonal elements of the FIM describe the information content associated with each parameter individually, while the off-diagonal elements encode covariances and parameter degeneracies. Since our interest is understanding how undersampling leads to a degeneracy between the amplitude and width, we restrict attention to the $(A,\sigma)$ subspace of the Fisher matrix. The relevant elements are

\begin{align}
    \mathcal{I}_{AA}
    &=
    \frac{1}{\sigma_{\mathrm{noise}}^2}
    \sum_i
    \left(\frac{\partial f_i}{\partial A}\right)^2
    =
    \frac{1}{\sigma_{\mathrm{noise}}^2}
    \sum_i
    \exp\!\left[-\frac{(v_i-\mu)^2}{\sigma^2}\right],
    \\[6pt]
    \mathcal{I}_{A\sigma}
    &=
    \frac{1}{\sigma_{\mathrm{noise}}^2}
    \sum_i
    \frac{\partial f_i}{\partial A}
    \frac{\partial f_i}{\partial \sigma},
    \\[6pt]
    \mathcal{I}_{\sigma\sigma}
    &=
    \frac{1}{\sigma_{\mathrm{noise}}^2}
    \sum_i
    \left(\frac{\partial f_i}{\partial \sigma}\right)^2.
\end{align}

\noindent
The off-diagonal term $\mathcal{I}_{A\sigma}$ quantifies the degree of degeneracy between the amplitude and width parameters. When the line profile is well sampled, this covariance is small, and the uncertainty on $A$ is well approximated by $\mathcal{I}_{AA}^{-1/2}$. However, as the sampling becomes coarser, the amplitude and width become increasingly correlated, inflating the marginal uncertainty on $A$ beyond the Fisher prediction based on $\mathcal{I}_{AA}$ alone. When the grid spacing $\Delta v$ is small compared to $\sigma$, the sampling is effectively continuous, and the off-diagonal term $\mathcal{I}_{A\sigma}$ has little impact. However, as $\Delta v$ increases, the line core becomes under-resolved, the sum over $i$ is truncated, and the correlation coefficient

\begin{equation}
    \rho_{A\sigma} = \frac{\mathcal{I}_{A\sigma}}{\sqrt{\mathcal{I}_{AA}\mathcal{I}_{\sigma\sigma}}}
\end{equation}

\noindent
approaches a positive value, reducing the effective Fisher information for $A$. The marginal variance on $A$ then becomes

\begin{equation}
    \mathrm{Var}(A) = (\mathcal{I}^{-1})_{AA} =
    \frac{1}{\mathcal{I}_{AA}(1-\rho_{A\sigma}^2)}.
\end{equation}

\noindent
Consequently, the standard deviation is inflated by a factor

\begin{equation}
    f_{\mathrm{cov}} = \frac{1}{\sqrt{1-\rho_{A\sigma}^2}}.
\end{equation}

\noindent
Substituting this relation yields an empirical inflation model:

\begin{equation}
    f_{\mathrm{cov}}(\Delta v) =
    1 + (f_{\mathrm{max}} - 1)
    \bigg[1 - \exp\!\bigg(-\frac{\Delta v^2}{\sigma^2}\bigg)\bigg],
\end{equation}

This function reproduces the observed behaviour in our numerical experiment: negligible inflation for well-sampled profiles, and a steady increase in $\sigma_A$ once the sampling step becomes a significant fraction of $\sigma$. The analytic curve including this covariance term (Fig.~\ref{fig:amp_unc_vs_step}, purple line) accurately tracks the behaviour of the direct-fit uncertainties.

\setcounter{section}{1}
\setcounter{table}{0}
\setcounter{figure}{0}

\begin{table*}[htbp]
\centering
\caption{Relative weights assigned to each observing night for each detected species.}
\label{tab:species_weights}
\scriptsize
\setlength{\tabcolsep}{3pt}
\renewcommand{\arraystretch}{1.05}
\resizebox{\textwidth}{!}{
\begin{tabular}{l
S S S S S
S S S S S
S S S}
\hline\hline
\multicolumn{1}{l}{Species} &
\multicolumn{1}{c}{N1} &
\multicolumn{1}{c}{N2} &
\multicolumn{1}{c}{N3} &
\multicolumn{1}{c}{N4} &
\multicolumn{1}{c}{N5} &
\multicolumn{1}{c}{N6} &
\multicolumn{1}{c}{N7} &
\multicolumn{1}{c}{N8} &
\multicolumn{1}{c}{N9} &
\multicolumn{1}{c}{N10} &
\multicolumn{1}{c}{N11} &
\multicolumn{1}{c}{N12} &
\multicolumn{1}{c}{N13} \\

\hline
HI   &0.1726&0.1318&0.0567&0.0338&0.1455&0.0675&0.0319&0.0196&0.0040&0.0026&0.0381&0.2383&0.0578 \\

CI   &0.0076&0.0171&0.0616&0.0789&0.0942&0.0178&0.0382&0.0207&0.0118&0.0040&0.0057&0.4664&0.1761 \\
OI   &0.0173&0.0096&0.0275&0.0189&0.0730&0.0214&0.0189&0.0301&0.0280&0.0102&0.0043&0.5613&0.1796 \\

NaI  &0.0872&0.0882&0.0374&0.0233&0.1046&0.0436&0.0307&0.0116&0.0068&0.0104&0.0462&0.4373&0.0727 \\
MgI  &0.1331&0.0924&0.0158&0.0109&0.0240&0.0457&0.0318&0.0247&0.0063&0.0018&0.0645&0.4240&0.1251 \\
SiI  &0.0748&0.0069&0.0068&0.0148&0.0626&0.1644&0.0597&0.0448&0.0187&0.0205&0.1165&0.2674&0.1423 \\
KI   &0.0771&0.0307&0.1923&0.0843&0.1631&0.0211&0.0194&0.0722&0.0779&0.0502&0.0532&0.0683&0.0902 \\

CaI  &0.1536&0.0563&0.0054&0.0329&0.0425&0.0673&0.0474&0.0048&0.0194&0.0053&0.1189&0.3316&0.1147 \\
CaII &0.0674&0.0052&0.0381&0.0222&0.2419&0.0244&0.0018&0.0102&0.0027&0.0016&0.0173&0.4297&0.1375 \\

ScII  &0.1723&0.0570&0.0255&0.0208&0.0089&0.1939&0.0360&0.0575&0.0209&0.0044&0.2805&0.0825&0.0398 \\

TiI  &0.1437&0.0430&0.0204&0.0000&0.0089&0.0797&0.0093&0.0053&0.0264&0.0089&0.3897&0.1694&0.0952 \\
TiII &0.1865&0.0981&0.0005&0.0037&0.0021&0.0948&0.0414&0.0071&0.0085&0.0052&0.4724&0.0626&0.0172 \\

VI   &0.2199&0.0776&0.0215&0.0863&0.0191&0.0597&0.1355&0.0530&0.0252&0.0112&0.1254&0.0940&0.0716 \\
VII  &0.1398&0.1364&0.0119&0.0127&0.0164&0.1810&0.0491&0.0099&0.0412&0.0300&0.3036&0.0523&0.0156 \\

CrI  &0.1588&0.0496&0.0149&0.0178&0.0091&0.0884&0.0369&0.0397&0.0027&0.0205&0.1184&0.3768&0.0664 \\
CrII &0.1313&0.0807&0.0094&0.0016&0.0053&0.0520&0.0179&0.0160&0.0025&0.0064&0.5341&0.0815&0.0611 \\

MnI  &0.1408&0.1318&0.0449&0.0146&0.0103&0.2042&0.0612&0.0189&0.0274&0.0069&0.0290&0.2079&0.1022 \\

FeI  &0.1490&0.0963&0.0149&0.0175&0.0459&0.0281&0.0292&0.0097&0.0014&0.0083&0.1006&0.3648&0.1344 \\
FeII &0.1286&0.1167&0.0053&0.0093&0.0249&0.0773&0.0281&0.0203&0.0043&0.0067&0.1323&0.3192&0.1268 \\

CoI  &0.1299&0.0260&0.0459&0.0285&0.0239&0.1095&0.0528&0.0139&0.0132&0.0206&0.4497&0.0474&0.0386 \\

NiI  &0.1156&0.0906&0.0403&0.0548&0.1161&0.0737&0.0185&0.0563&0.0805&0.0128&0.2409&0.0624&0.0375 \\
NiII &0.0297&0.0358&0.0320&0.0768&0.0198&0.0375&0.0792&0.0799&0.0795&0.0434&0.3567&0.0991&0.0306 \\

SrII &0.3740&0.0422&0.0178&0.0125&0.0134&0.1851&0.0804&0.0433&0.0350&0.0708&0.1052&0.0057&0.0147 \\

YII  &0.1111&0.0913&0.0355&0.0151&0.0245&0.1017&0.0574&0.0335&0.0254&0.0188&0.1380&0.3189&0.0288 \\

ZrII &0.1794&0.0280&0.0061&0.0232&0.0339&0.1095&0.0490&0.0326&0.0261&0.0138&0.4279&0.0259&0.0447 \\

BaII &0.0934&0.0998&0.0315&0.0463&0.0450&0.1076&0.0267&0.0099&0.0209&0.0194&0.1354&0.2721&0.0922 \\

LaII &0.1174&0.0641&0.0596&0.0517&0.0456&0.1682&0.0354&0.0431&0.0096&0.0060&0.0827&0.1510&0.1654 \\
PrII &0.1742&0.1050&0.0537&0.0563&0.0452&0.1292&0.0912&0.0664&0.0693&0.0384&0.1019&0.0323&0.0367 \\
NdII &0.0872&0.1124&0.0127&0.0123&0.0609&0.0556&0.0421&0.0995&0.0138&0.0443&0.1099&0.2645&0.0847 \\

\hline
\end{tabular}}
\tablefoot{
Species are ordered by atomic number and ionisation state.
Observing nights correspond to:
N1: HARPS-N (Night~1),
N2: HARPS-N (Night~2),
N3: CARMENES VIS (Night~1),
N4: CARMENES VIS (Night~2),
N5: CARMENES VIS (Night~3),
N6: FIES (Night~1),
N7: FIES (Night~2),
N8: FOCES (Night~1),
N9: FOCES (Night~2),
N10: FOCES (Night~3),
N11: HIRES (Night~1),
N12: MAROON-X (Night~1),
N13: MAROON-X (Night~2).
}
\end{table*}
\begin{table*}[htbp]
  \centering
  \small
  \setlength{\tabcolsep}{10pt}
  \renewcommand{\arraystretch}{1.2}
  \caption{Fraction of masked pixels identified for each observing night.}
  \label{tab:badpix_stats}
  \begin{tabular}{l c c c c c}
    \hline\hline
    Instrument & Night & Total pixels & Broad masks & Isolated pixels & Total masked \\
     & & & [\%] & [\%] & [\%] \\
    \hline
    HARPS-N    & N1  & $1.38\times10^{7}$ & 0.37 & 0.00 & 0.37 \\
    HARPS-N    & N2  & $1.30\times10^{7}$ & 0.98 & 0.08 & 1.06 \\
    CARMENES   & N3  & $1.35\times10^{7}$ & 17.13 & 0.16 & 17.29 \\
    CARMENES   & N4  & $1.25\times10^{7}$ & 16.77 & 2.28 & 19.05 \\
    CARMENES   & N5  & $3.50\times10^{7}$ & 15.94 & 0.09 & 16.02 \\
    FIES       & N6  & $6.76\times10^{6}$ & 4.59 & 0.09 & 4.68 \\
    FIES       & N7  & $7.32\times10^{6}$ & 3.64 & 0.17 & 3.81 \\
    FOCES      & N8  & $7.05\times10^{6}$ & 5.76 & 0.68 & 6.44 \\
    FOCES      & N9  & $6.02\times10^{6}$ & 7.10 & 0.70 & 7.80 \\
    FOCES      & N10 & $6.71\times10^{6}$ & 6.52 & 0.85 & 7.37 \\
    HIRES      & N11 & $9.52\times10^{6}$ & 18.34 & 0.78 & 19.13 \\
    MAROON-X   & N12 & $1.43\times10^{7}$ & 4.51 & 0.06 & 4.58 \\
    MAROON-X   & N13 & $1.53\times10^{7}$ & 5.39 & 0.28 & 5.66 \\
    \hline
  \end{tabular}
  \tablefoot{
    Broad masks correspond to manually masked detector columns or wavelength regions affected by telluric residuals, low throughput, or order-level artefacts.
    Isolated pixels correspond to NaNs flagged using a median absolute deviation (MAD) filter.
    Percentages are given relative to the total number of pixels for each observing night.
  }
\end{table*}
\begin{figure*}
    \centering
    \includegraphics[width=\textwidth,keepaspectratio]{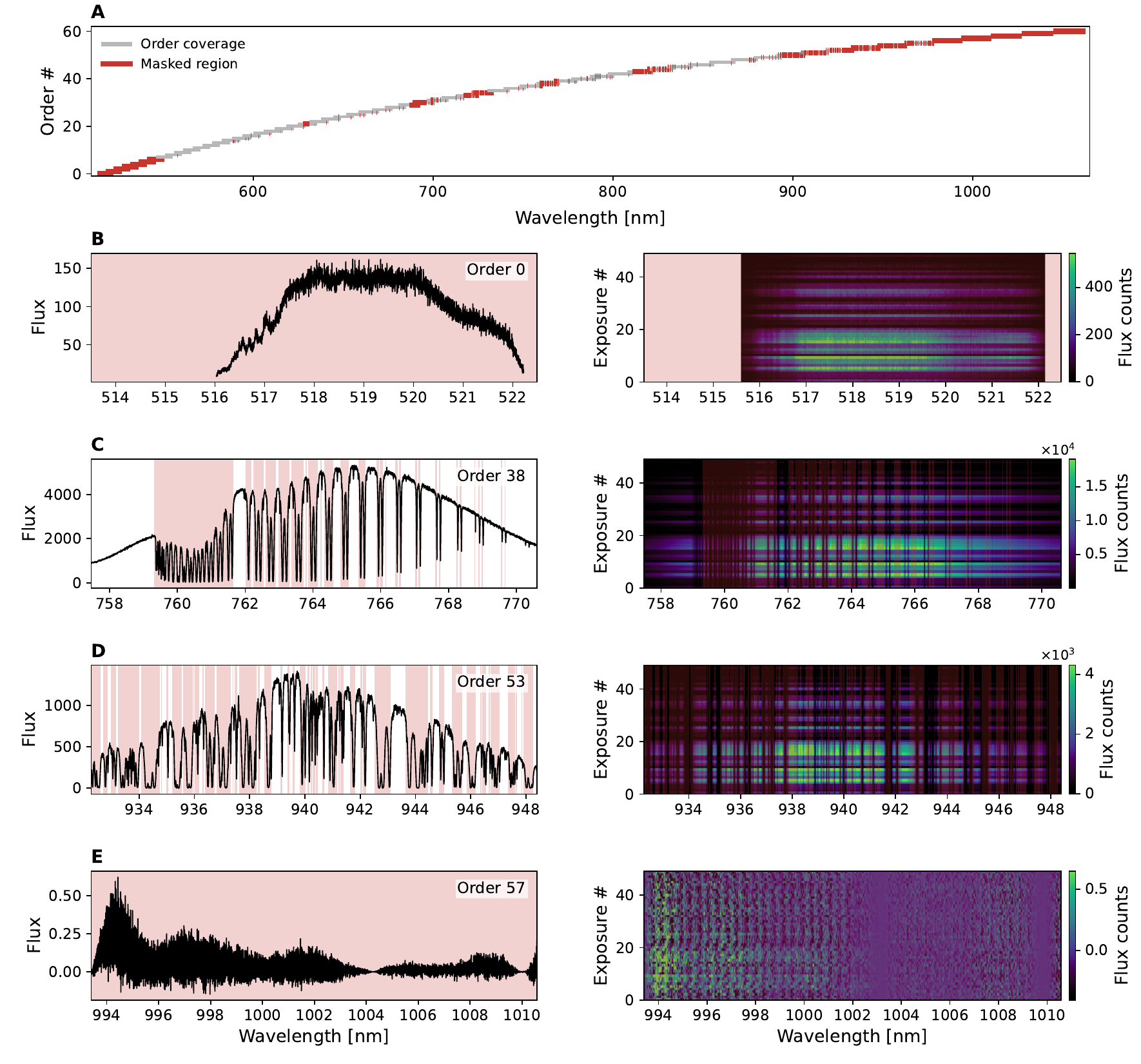}
    \caption{CARMENES masking diagnostic for the representative high masked-pixel fraction case. Panel A shows the wavelength coverage of each CARMENES VIS order for the second CARMENES night, which has the largest CARMENES masked fraction in Table~\ref{tab:badpix_stats}. Panels B--E show representative order-level examples; the left panels show the average extracted flux counts for the selected order, while the right panels show the corresponding exposure-by-exposure flux counts.}
    \label{fig:carmenes_mask_diagnostic}
\end{figure*}
\begin{figure*}
    \centering
    \includegraphics[width=\textwidth,keepaspectratio]{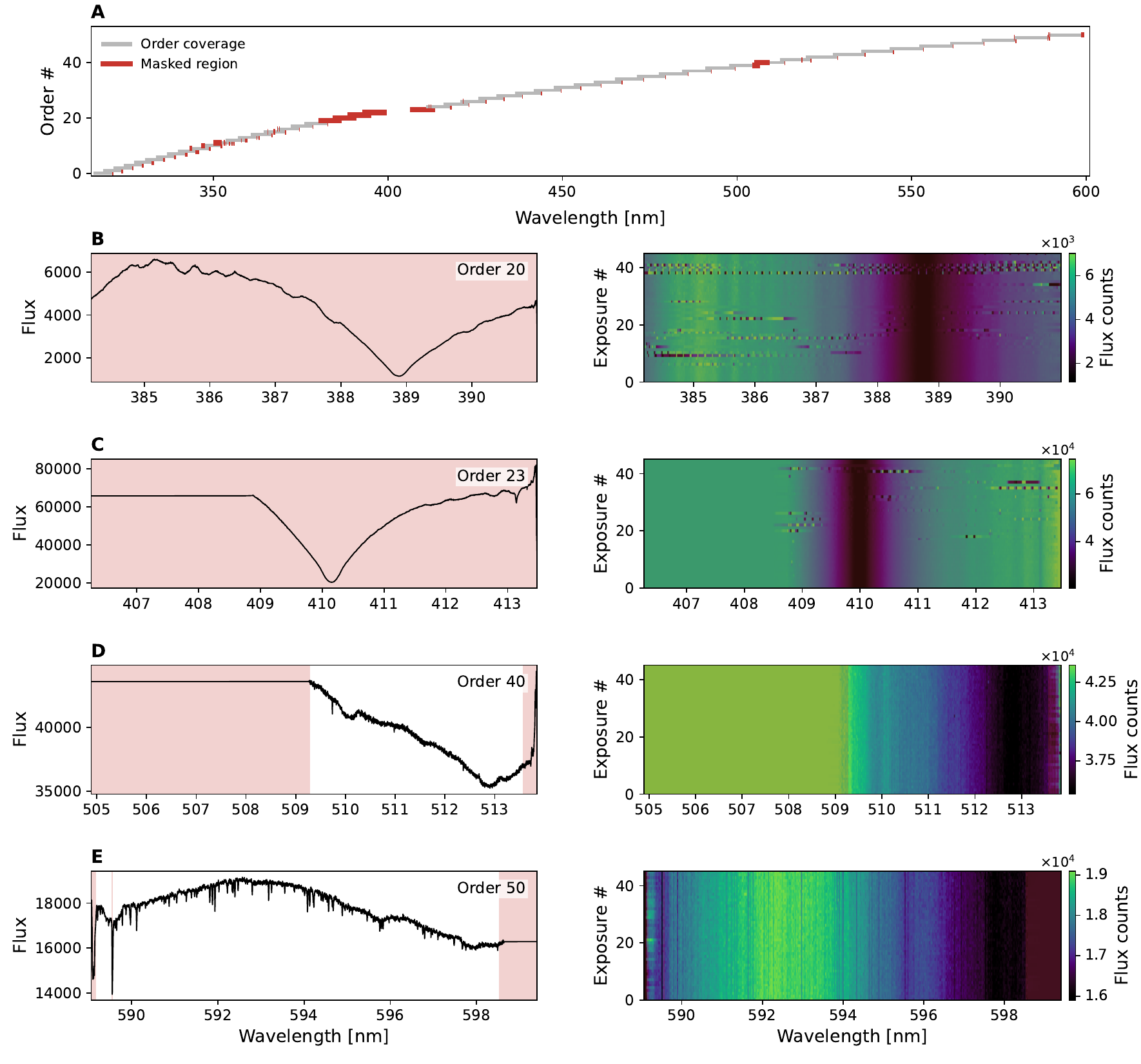}
    \caption{HIRES masking diagnostic and representative examples of the order-level structures discussed in Sect.~\ref{sect:keckhires}. Panel A shows the wavelength coverage of each HIRES order, with masked regions highlighted in red. Panels B--E show representative order-level examples; the left panels show the average extracted flux counts for the selected order, while the right panels show the corresponding exposure-by-exposure flux counts.}
    \label{fig:hires_mask_diagnostic}
\end{figure*}
\setcounter{section}{2}
\setcounter{table}{0}
\setcounter{figure}{0}
\begin{table*}[htbp]
\centering
\small
\caption{Wavelength intervals used for telluric fitting for each spectrograph.}
\label{tab:telluric_regions}
\begin{tabular}{ll}
\hline\hline
Spectrograph & Inclusion regions \\
 & ($\mu$m) \\
\hline
HARPS-N   & 0.5908–0.5939, 0.5954–0.5984, 0.6288–0.6313, 0.6482–0.6513, 0.6880–0.6887, 0.6896–0.6901, 0.6904–0.6909 \\
CARMENES  & 0.5908–0.5924, 0.6295–0.6311, 0.6464–0.6474, 0.6493–0.6512, 0.6883–0.6919, 0.6947–0.6965, 0.7014–0.7036, \\
          & 0.7215–0.7232, 0.7340–0.7388, 0.7670–0.7695, 0.7702–0.7709, 0.7907–0.7928, 0.8052–0.8090, 0.8222–0.8252, \\
          & 0.8347–0.8426, 0.8925–0.8962, 0.9086–0.9125, 0.9709–0.9722, 0.9743–0.9755 \\
FIES      & 0.5904–0.6000, 0.6288–0.6315, 0.6469–0.6495, 0.6883–0.6922, 0.6985–0.6994, 0.7036–0.7051, 0.7162–0.7192, \\
          & 0.7676–0.7703, 0.8256–0.8287, 0.8922–0.8971 \\
FOCES     & 0.5935–0.5963, 0.6275–0.6298, 0.6472–0.6501, 0.6857–0.6884, 0.6896–0.6913, 0.7052–0.7063, 0.7156–0.7217, \\
          & 0.7217–0.7232, 0.7287–0.7353, 0.7655–0.7669, 0.7676–0.7716, 0.8251–0.8327 \\
HIRES     & 0.5910–0.5923 \\
MAROON-X  & 0.6300–0.6313, 0.6885–0.6897, 0.6948–0.6961, 0.7288–0.7307, 0.7654–0.7694, 0.8269–0.8312, 0.8956–0.8981 \\
\hline
\end{tabular}
\tablefoot{
Inclusion regions correspond to wavelength intervals dominated by telluric absorption and used for fitting with \texttt{Molecfit}.
}
\end{table*}
\setcounter{section}{3}
\setcounter{table}{0}
\setcounter{figure}{0}
\begin{figure*}[htbp]
    \centering
    \includegraphics[width=\textwidth,keepaspectratio]{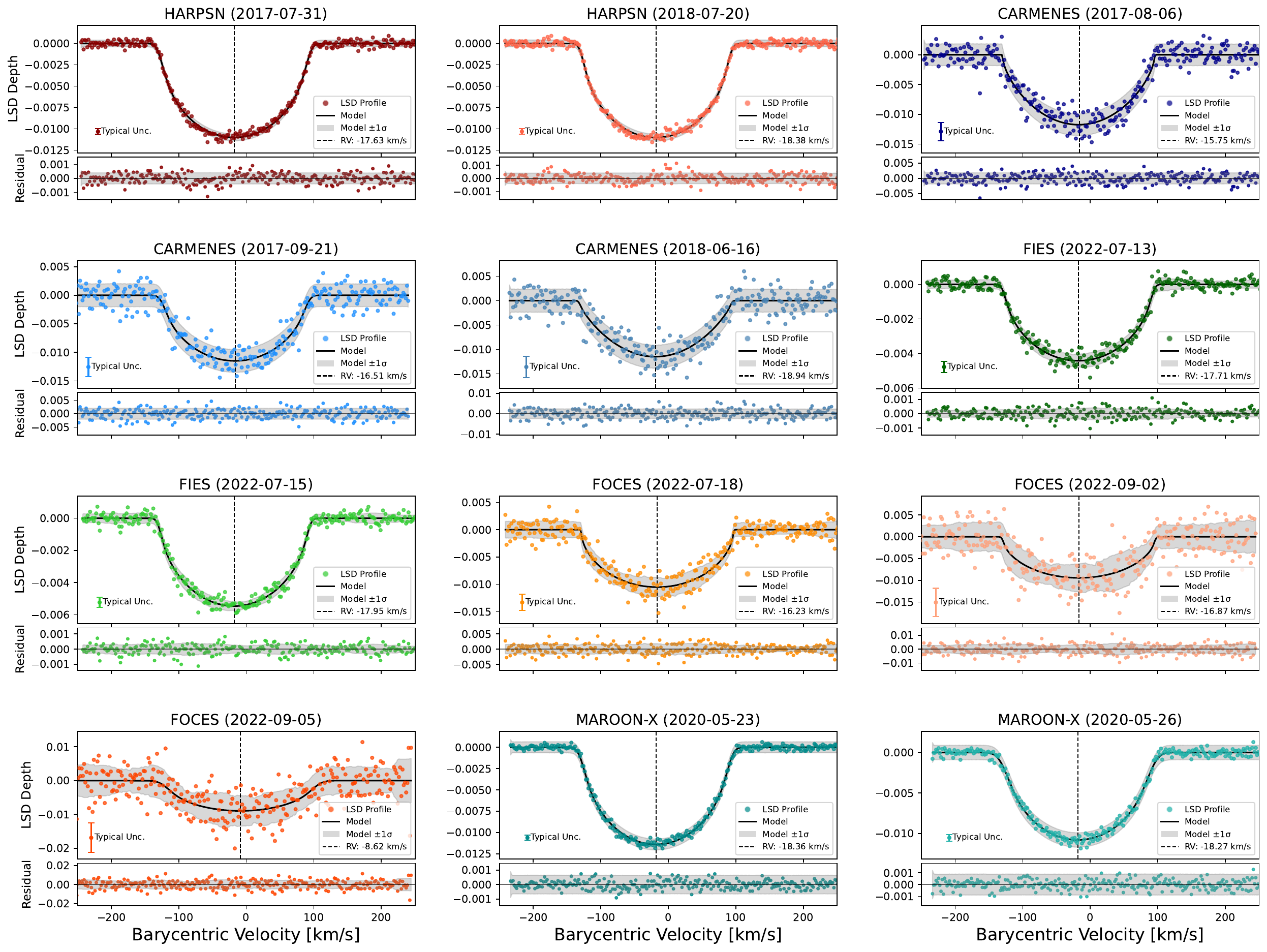}
    \caption{LSD profile fits and residuals for the first out-of-transit exposure of each observing night. Each panel shows the least-squares deconvolved (LSD) profile (coloured points) alongside the best-fit rotational profile (black line) and its 1$\sigma$ model uncertainty (grey shaded region). A representative uncertainty bar (labelled ``Typical Unc.'') is shown in the bottom-left of each panel, representing the root-mean-square of the residuals for that exposure. Residuals between the observed and model LSD profiles are shown beneath each panel, illustrating the goodness of fit and absence of systematic structure. Colours correspond to the observing instrument, following the same scheme used throughout the manuscript.
    }
    \label{fig:EX_first_exposures_grid}
\end{figure*}
\begin{table}[htbp]
  \centering
  \scriptsize
  \setlength{\tabcolsep}{10pt}
  \renewcommand{\arraystretch}{1.2}
  \caption{Systemic velocity statistics grouped by spectrograph.}
  \label{tab:vsys_by_spectrograph}
  \begin{tabularx}{\columnwidth}{>{\centering\arraybackslash}X
                                   >{\centering\arraybackslash}X
                                   >{\centering\arraybackslash}X
                                   >{\centering\arraybackslash}X}
    \hline\hline
    Spectrograph & Mean & Std. dev. & Std. error \\
     & [km\,s$^{-1}$] & [km\,s$^{-1}$] & [km\,s$^{-1}$] \\
    \hline
    HARPSN     & $-18.026$ & $0.388$ & $0.057$ \\
    CARMENES   & $-17.725$ & $2.353$ & $0.409$ \\
    FIES       & $-18.076$ & $0.877$ & $0.172$ \\
    FOCES      & $-18.088$ & $3.863$ & $0.822$ \\
    MAROON-X   & $-18.011$ & $0.353$ & $0.069$ \\
    \hline
  \end{tabularx}
  \tablefoot{
    Values correspond to the mean, standard deviation, and error on the mean of the systemic velocities measured in this work for each spectrograph.
  }
\end{table}
\setcounter{section}{4}
\setcounter{table}{0}
\setcounter{figure}{0}
\begin{table}[htbp]
  \centering
  \footnotesize
  \setlength{\tabcolsep}{5pt}
  \renewcommand{\arraystretch}{1.0}
  \caption{Mahalanobis $z$-scores and one-sided $p$-values for detected species.}
  \label{tab:zscores_species}
  \begin{tabularx}{\columnwidth}{X c c | X c c}
    \hline\hline
    Species & $z$-score & $p$ (1-sided) & Species & $z$-score & $p$ (1-sided) \\
    \hline
    Fe\,\textsc{ii}  & 29.65 & 1.47e-193 & Ni\,\textsc{ii}  &  8.62 & 3.26e-18 \\
    Fe\,\textsc{i}   & 18.78 & 5.79e-79 & Co\,\textsc{i}   &  8.46 & 1.31e-17 \\
    Ti\,\textsc{ii}  & 16.49 & 2.21e-61 & Ni\,\textsc{i}   &  8.24 & 8.72e-17 \\
    Ca\,\textsc{ii}  & 13.09 & 1.79e-39 & V\,\textsc{i}    &  8.18 & 1.38e-16 \\
    H\,\textsc{i}    & 12.00 & 1.69e-33 & K\,\textsc{i}    &  8.10 & 2.72e-16 \\
    O\,\textsc{i}    & 11.44 & 1.40e-30 & Zr\,\textsc{ii}  &  8.01 & 5.78e-16 \\
    Cr\,\textsc{ii}  & 11.24 & 1.30e-29 & C\,\textsc{i}    &  7.97 & 7.77e-16 \\
    Na\,\textsc{i}   & 10.91 & 5.10e-28 & Ti\,\textsc{i}   &  7.77 & 3.78e-15 \\
    Sc\,\textsc{ii}  & 10.80 & 1.77e-27 & La\,\textsc{ii}  &  7.54 & 2.32e-14 \\
    Mg\,\textsc{i}   & 10.29 & 3.76e-25 & Mn\,\textsc{i}   &  7.47 & 3.98e-14 \\
    V\,\textsc{ii}   & 10.23 & 7.36e-25 & Ba\,\textsc{ii}  &  7.24 & 2.30e-13 \\
    Nd\,\textsc{ii}  &  9.94 & 1.40e-23 & Si\,\textsc{i}   &  6.99 & 1.33e-12 \\
    Ca\,\textsc{i}   &  9.73 & 1.17e-22 & Pr\,\textsc{ii}  &  6.93 & 2.06e-12 \\
    Cr\,\textsc{i}   &  8.99 & 1.27e-19 & Sr\,\textsc{ii}  &  6.89 & 2.86e-12 \\
    Y\,\textsc{ii}   &  8.99 & 1.28e-19 & \multicolumn{1}{c}{--} & \multicolumn{1}{c}{--} & \multicolumn{1}{c}{--} \\
    \hline
  \end{tabularx}
  \tablefoot{
    Species are sorted by decreasing $z$-score, calculated from the 2D joint $(A_0, A_1)$ bootstrap covariance matrix.
    The $z$-score corresponds to the detection significance under the negative-amplitude (absorption) hypothesis.
  }
\end{table}
\begin{figure}[htbp]
  \centering
  \includegraphics[width=\columnwidth,height=0.68\textheight,keepaspectratio]{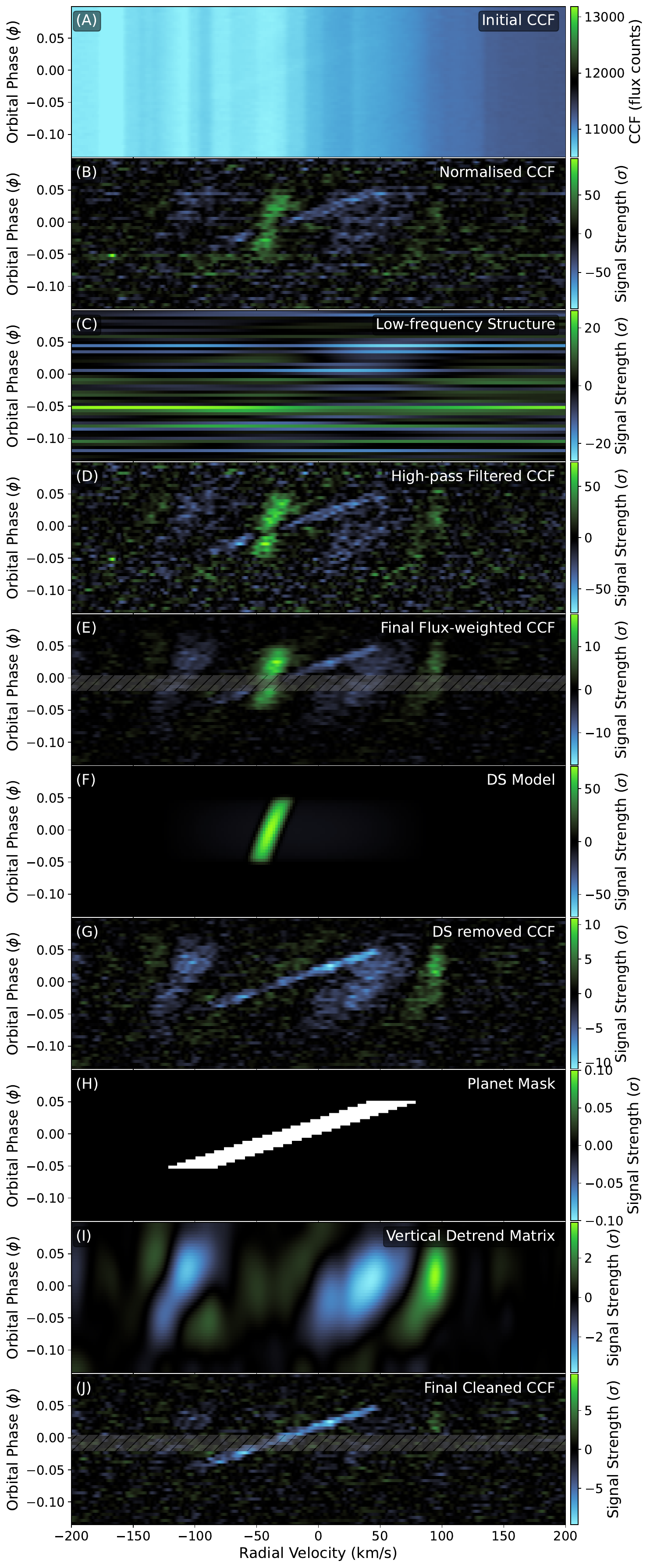}
  \caption{A step-by-step illustration of the CCF cleaning pipeline applied to one night of HARPS-N observations of KELT-9b. Panels (A)--(J) show: (A) initial raw CCFs; (B) normalised CCFs; (C) low-frequency structure removed by the high-pass filter; (D) high-pass--filtered CCFs; (E) final flux-weighted CCFs; (F) \texttt{StarRotator} Doppler-shadow model; (G) \texttt{StarRotator}-subtracted CCFs; (H) planetary mask; (I) vertical detrending matrix; and (J) final cleaned CCFs. Panels (B)--(J) are shown in units of $\sigma$, while Panel (A) is shown in flux counts. The shaded gray regions in Panels~E and J highlight the excluded planet--Rossiter--McLaughlin overlap phase interval ($-0.015 \le \phi \le 0.001$). Absorption is shown in blue and emission in green.}
  \label{fig:ccf_cleaning}
\end{figure}
\begin{figure}[htbp]
  \centering
  \includegraphics[width=\columnwidth,height=0.75\textheight,keepaspectratio]{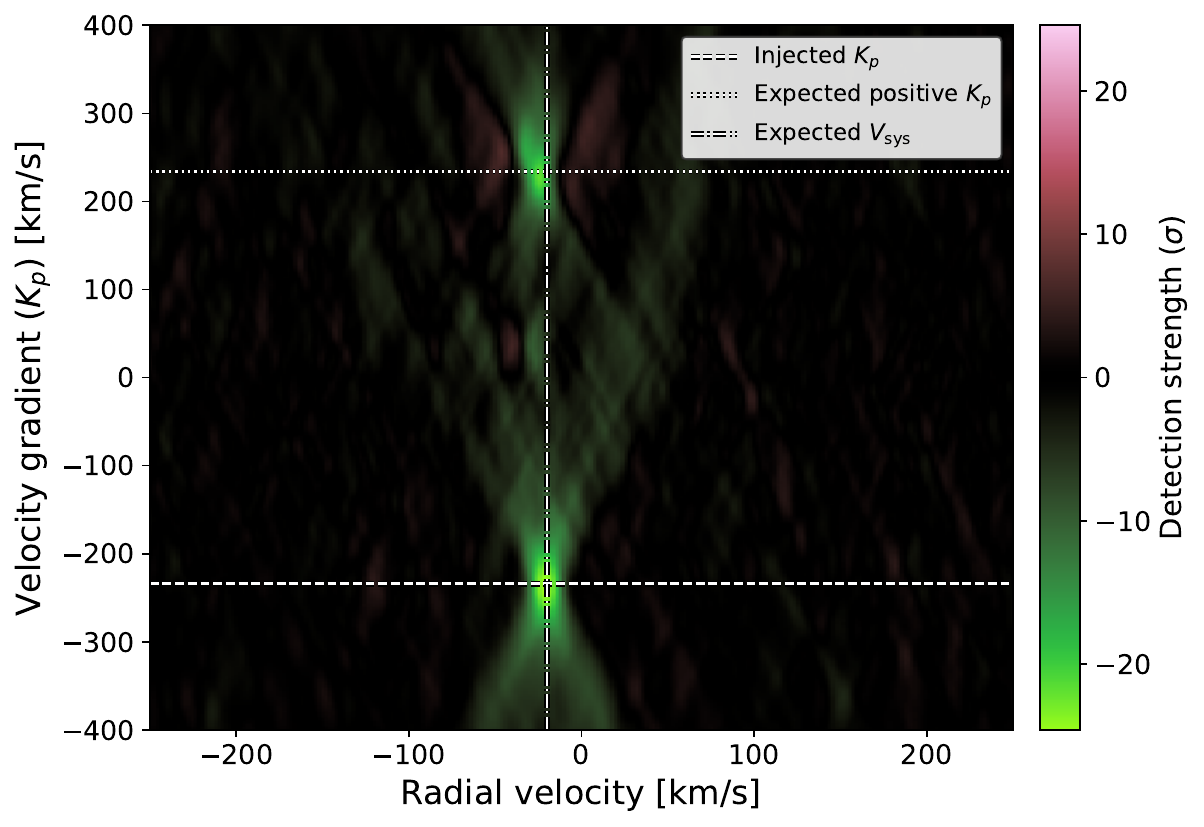}
  \caption{Negative-$K_p$ model-injection control for Fe\,I. The dashed, dotted, and dash-dotted lines mark the injected negative-$K_p$ track ($K_p=-234$~km\,s$^{-1}$), the expected positive-$K_p$ track, and the systemic velocity, respectively.}
  \label{fig:negative_kp_injection_control}
\end{figure}
\begin{table}[htbp]
  \centering
  \footnotesize
  \setlength{\tabcolsep}{6pt}
  \renewcommand{\arraystretch}{1.2}
  \caption{Comparison between a direct log-likelihood/MCMC fit and the adopted bootstrap framework for Fe\,II using the first HARPS-N night.}
  \label{tab:feii_loglikelihood_bootstrap}
  \begin{tabularx}{\columnwidth}{X c c}
    \hline\hline
    Parameter & Direct MCMC & Bootstrap \\
    \hline
    $K_p$ [km\,s$^{-1}$] & $237.9 \pm 0.3$ & $240.4 \pm 2.9$ \\
    $K_{p,1}$ [km\,s$^{-1}$] & $2.0 \pm 1.8$ & $24.2 \pm 16.8$ \\
    $V_{\rm sys}$ [km\,s$^{-1}$] & $-26.65 \pm 0.07$ & $-27.53 \pm 0.58$ \\
    $\sigma_0$ [km\,s$^{-1}$] & $8.37 \pm 0.05$ & $8.32 \pm 0.40$ \\
    $\sigma_1$ [km\,s$^{-1}$] & $-2.41 \pm 0.24$ & $-2.92 \pm 2.02$ \\
    $z$-score & $194.4$ & $19.6$ \\
    \hline
  \end{tabularx}
  \tablefoot{Columns compare the direct MCMC and bootstrap parameter estimates for the same first-HARPS-N Fe\,II CCF under the 7-parameter Mandel \& Agol model. The $z$-score row gives the Mahalanobis detection significance.}
\end{table}
\begin{table}[htbp]
  \centering
  \footnotesize
  \renewcommand{\arraystretch}{1.2}
  \caption{Species-specific systemic velocity offsets ($V_{0,\mathrm{sys}}$) derived from the bootstrap fits.}
  \label{tab:v0_species}
  \begin{tabularx}{\columnwidth}{X c | X c}
    \hline\hline
    Species & $V_{0,\mathrm{sys}}$ & Species & $V_{0,\mathrm{sys}}$ \\
            & [km\,s$^{-1}$] & & [km\,s$^{-1}$] \\
    \hline
    Fe\,\textsc{ii}  & $-26.33 \pm 0.33$ & Sr\,\textsc{ii}  & $-23.87 \pm 1.42$ \\
    Ti\,\textsc{ii}  & $-27.12 \pm 0.43$ & Mg\,\textsc{i}   & $-26.88 \pm 1.22$ \\
    H\,\textsc{i}    & $-24.35 \pm 1.63$ & Ni\,\textsc{i}   & $-26.35 \pm 1.41$ \\
    Cr\,\textsc{ii}  & $-27.57 \pm 0.71$ & Ca\,\textsc{i}   & $-27.24 \pm 1.25$ \\
    Ca\,\textsc{ii}  & $-26.65 \pm 1.10$ & La\,\textsc{ii}  & $-21.22 \pm 1.29$ \\
    Fe\,\textsc{i}   & $-26.91 \pm 0.62$ & K\,\textsc{i}    & $-25.23 \pm 1.30$ \\
    Cr\,\textsc{i}   & $-27.75 \pm 1.22$ & V\,\textsc{i}    & $-25.72 \pm 1.41$ \\
    O\,\textsc{i}    & $-26.11 \pm 0.89$ & Nd\,\textsc{ii}  & $-25.56 \pm 1.16$ \\
    V\,\textsc{ii}   & $-23.38 \pm 0.94$ & Pr\,\textsc{ii}  & $-25.46 \pm 1.33$ \\
    Na\,\textsc{i}   & $-25.53 \pm 0.92$ & Y\,\textsc{ii}   & $-25.41 \pm 1.39$ \\
    Sc\,\textsc{ii}  & $-22.40 \pm 0.98$ & C\,\textsc{i}    & $-27.16 \pm 1.44$ \\
    Zr\,\textsc{ii}  & $-24.84 \pm 1.45$ & Ti\,\textsc{i}   & $-25.24 \pm 1.52$ \\
    Ni\,\textsc{ii}  & $-24.40 \pm 1.72$ & Si\,\textsc{i}   & $-21.67 \pm 1.64$ \\
    Ba\,\textsc{ii}  & $-25.39 \pm 1.50$ & Mn\,\textsc{i}   & $-24.63 \pm 1.53$ \\
    Co\,\textsc{i}   & $-25.36 \pm 1.46$ & \multicolumn{1}{c}{--} & \multicolumn{1}{c}{--} \\
    \hline
  \end{tabularx}
  \tablefoot{
    Values correspond to the median (50th percentile) and standard deviation across the 100 combined bootstrap realisations.
  }
\end{table}
\begin{table}[htbp]
  \centering
  \footnotesize
  \renewcommand{\arraystretch}{1.2}
  \caption{Likely interstellar medium (ISM) absorption features identified and masked during visual inspection.}
  \label{tab:ism_features}
  \begin{tabularx}{\columnwidth}{>{\centering\arraybackslash}X >{\centering\arraybackslash}X}
    \hline\hline
    Species & Air wavelength \\
            & (\AA) \\
    \hline
    Na\,\textsc{i} D$_2$ & 5889.95 \\
    Na\,\textsc{i} D$_1$ & 5895.92 \\
    \hline
  \end{tabularx}
  \tablefoot{
    Wavelengths are given in air. These narrow features were consistently observed across all instruments and were masked to prevent contamination of the planetary signal.
  }
\end{table}
\begin{figure*}[htbp]
  \centering
  \includegraphics[width=\textwidth,height=0.75\textheight,keepaspectratio]{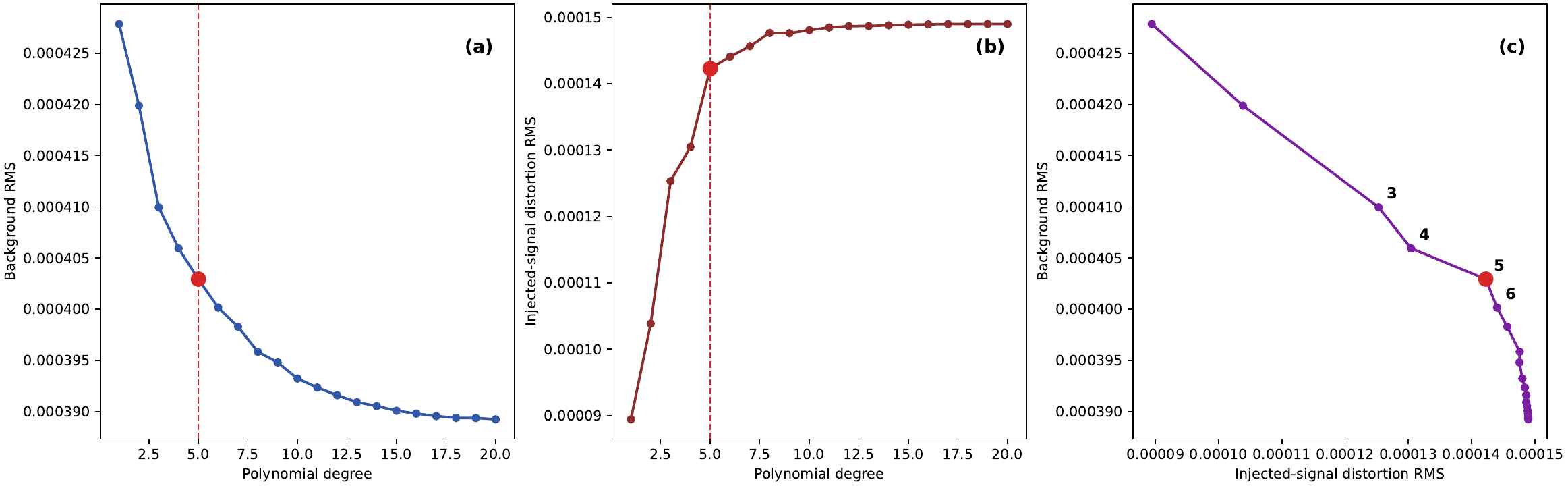}
  \caption{Polynomial-order sensitivity diagnostics. Panels show: (a) residual background RMS outside the planetary trail ($100 \le |v| \le 300\mathrm{\,km\,s^{-1}}$) as a function of polynomial degree; (b) RMS distortion of the processed injected forward model within the planet search region ($|v| \le 50\mathrm{\,km\,s^{-1}}$); and (c) joint background--distortion trade-off. The adopted degree-5 point is highlighted as the optimal balance point where the rapid background-RMS reduction has flattened while injected-signal distortion continues to increase.}
  \label{fig:r4_07_order_selection}
\end{figure*}
\begin{table}[htbp]
  \centering
  \scriptsize
  \renewcommand{\arraystretch}{1.15}
  \caption{Polynomial-order detrending diagnostics for degrees 1--10.}
  \label{tab:r4_07_order_metrics}
  \begin{tabular}{c c c}
    \hline\hline
    Degree & Background RMS ($\times 10^{-4}$) & Injected distortion RMS ($\times 10^{-4}$) \\
    \hline
    1 & 4.279 & 0.894 \\
    2 & 4.199 & 1.038 \\
    3 & 4.099 & 1.253 \\
    4 & 4.059 & 1.304 \\
    5 & 4.029 & 1.423 \\
    6 & 4.002 & 1.440 \\
    7 & 3.983 & 1.456 \\
    8 & 3.958 & 1.476 \\
    9 & 3.948 & 1.476 \\
    10 & 3.932 & 1.480 \\
    \hline
  \end{tabular}
  \tablefoot{The background RMS is measured outside the planetary-search region ($100 \le |v| \le 300\mathrm{\,km\,s^{-1}}$). The injected-signal distortion is the RMS difference between the true injected model and the recovered signal within the planet velocity region ($|v| \le 50\mathrm{\,km\,s^{-1}}$). Degree 5 is the adopted baseline.}
\end{table}
\end{appendices}

\end{document}